\documentclass[aps,pra,reprint,twocolum,amsmath,superscriptaddress,amssymb,showpacs,nofootinbib]{revtex4-2}
\usepackage{amsmath,braket,amssymb}
\usepackage[final]{graphicx}
\usepackage{subfigure}
\usepackage[english]{babel}
\usepackage[utf8]{inputenc}
\usepackage{mathtools}
\usepackage{empheq}
\usepackage{tensor}
\usepackage{cancel}
\usepackage{ytableau}
\usepackage{comment}
\usepackage{enumitem}
\usepackage{slashed}
\usepackage{graphicx}
\usepackage{xcolor}

\usepackage{bm}
\usepackage[colorlinks=true,citecolor=blue,linkcolor=blue,urlcolor=blue]{hyperref}
\usepackage{dsfont}
\usepackage{changepage}
\usepackage{array}
\usepackage{soul}
\usepackage[capitalise]{cleveref}
\crefname{section}{Sec.}{Secs.}
\Crefname{section}{Sec.}{Secs.}
\usepackage{xifthen}
\usepackage{xargs}

\newcommand{\NLSM}{$\rm NL\sigma M$}

\newcommand{\mJ}{\mathcal{J}}

\usepackage{bm}
\renewcommand{\vec}[1]{\boldsymbol{\mathbf{#1}}}
\renewcommand{\det}{\operatorname{det}}

\graphicspath{{./}{./figures/}}

\newcommand{\bit}{\begin{itemize}}
\newcommand{\eit}{\end{itemize}}

\newcommand{\f}{\frac}
\renewcommand{\>}{\right\rangle}
\newcommand{\<}{\left\langle}
\newcommand{\ba}{\begin{align}}
\newcommand{\ena}{\end{align}}
\newcommand{\be}{\begin{equation}}
\newcommand{\ee}{\end{equation}}
\newcommand{\bi}{\begin{itemize}}
\newcommand{\ei}{\end{itemize}}
\newcommand{\lf}{\left(}
\newcommand{\ri}{\right)}
\newcommand{\dd}{\mathrm{d}}

\newcommand{\Tr}{\operatorname{Tr}}
\newcommand{\tr}{\operatorname{tr}}

\DeclareMathAlphabet{\mymathbb}{U}{BOONDOX-ds}{m}{n}

\renewcommand{\log}{\ln}

\newcommand{\Hs}{\mathcal{H}}

\usepackage{braket}
\usepackage[normalem]{ulem}

\begin{document}
\date{\today}

\newcommand{\bbra}[1]{\<\< #1 \right|\right.}
\newcommand{\kket}[1]{\left.\left| #1 \>\>}
\newcommand{\bbrakket}[1]{\< \Braket{#1} \>}
\newcommand{\pll}{\parallel}
\newcommand{\nn}{\nonumber}
\newcommand{\transp}{\text{transp.}}
\newcommand{\nor}{z_{J,H}}

\newcommand{\hL}{\hat{L}}
\newcommand{\hR}{\hat{R}}
\newcommand{\hQ}{\hat{Q}}

\newcommand{\MF}[1]{{\color{olive} #1}}

\title{Monitored fermions with conserved $\mathrm{U}(1)$ charge}

\author{Michele Fava}
\affiliation{Philippe Meyer Institute, Physics Department, \'{E}cole Normale Sup\'{e}rieure (ENS), Universit\'{e} PSL, 24 rue Lhomond, F-75231 Paris, France}

 \author{Lorenzo Piroli}
 \affiliation{Dipartimento di Fisica e Astronomia, Università di Bologna and INFN,
Sezione di Bologna, via Irnerio 46, I-40126 Bologna, Italy}

\author{Denis Bernard}
\affiliation{Laboratoire de Physique de l’\'Ecole Normale Sup\'erieure, CNRS, ENS \& Universit\'e PSL, Sorbonne Universit\'e, Universit\'e Paris Cit\'e, 75005 Paris, France}

\author{Adam Nahum}
\affiliation{Laboratoire de Physique de l’\'Ecole Normale Sup\'erieure, CNRS, ENS \& Universit\'e PSL, Sorbonne Universit\'e, Universit\'e Paris Cit\'e, 75005 Paris, France}

\begin{abstract}
We study measurement-induced phases of free fermion systems with U(1) symmetry.
Following a recent approach developed for Majorana chains, we derive a field theory description for the purity and bipartite entanglement at large space and time scales. We focus on a multi-flavor one-dimensional chain with random complex hoppings and continuous monitoring of the local fermion density. By means of the replica trick, and using the number of flavors as a large parameter controlling our approximations, we derive an effective field theory made up of a SU(N) non-linear sigma model (NL$\sigma$M) coupled to fluctuating hydrodynamics. Contrary to the case of non-interacting Majorana fermions, displaying no U(1) symmetry, we find that the bipartite entanglement entropy satisfies an area law for all monitoring rates, but with a nontrivial scaling of entanglement when the correlation length is large. We provide numerical evidence supporting our claims. We briefly show how imposing a reality condition on the hoppings can change the NL$\sigma$M and also discuss higher dimensional generalizations.
\end{abstract}

\maketitle

\section{Introduction}

Ongoing monitoring of a quantum many-body system, by an external observer, makes its dynamics stochastic as a result of quantum measurement back action and leads to new dynamical universality classes \cite{li2018quantum,skinner2019measurement}.
Formulating long-wavelength theories for these universality classes, in interacting systems, can be challenging. 
But in fact even for monitored \textit{free} fermions \cite{carollo2017fluctuating,Bernard_2018,cao2019entanglement}
 the classification of phases can be nontrivial \cite{nahum2020entanglement,alberton2021entanglement,jian2022criticality,alberton2021entanglement,turkeshi2021measurementinduced,buchold2021effective,sang2021measurement,kells2021topological,fidkowski2021dynamical,turkeshi2022enhanced,turkeshi2022entanglement,piccitto2022entanglement,muller2022measurement,turkeshi2022entanglementand,lucas2022generalized,coppola2022growth,gal2022volume,merritt2022entanglement,chen2020emergent,kells2021topological,paviglianiti2023multipartite,turkeshi2024density,leung2023theory,ravindranath2023free, piccitto2023entanglement,russomanno2023entanglement,piccitto2024impact,turkeshi2021measurementinduced, turkeshi2022entanglementand,gal2022volume,granet2022volume, jian2020measurement,bao2020theory,nahum2021measurement,fava2023nonlinear,jian2023measurement,poboiko2023theory}, 
 and has been controversial in the case where fermion number is conserved.

In monitored dynamics, the sequence of random measurement outcomes defines the ``quantum trajectory'' for the evolution of the state. We analyze quantum trajectories of  monitored {free} fermion systems with conserved  fermion number.
Fermions hop with a Hamiltonian ${H=\sum_{\<ij\>}J_{ij} {c_i}^{\hspace{-1.4pt}\dag} {c_j}^{\hspace{-1.4pt}\phantom{\dag}}}$,
while an observer makes repeated measurements of the local densities ${c_i}^{\hspace{-1.4pt}\dag} \hspace{-1pt}{c_i}^{\hspace{-1.4pt}\phantom{\dag}}$ \cite{carollo2017fluctuating,Bernard_2018,cao2019entanglement}.
The full dynamics is ``free'' in the sense that the measurements preserve the Gaussianity of the state in each quantum trajectory (i.e. for each sequence of measurement outcomes),
but it is nontrivial because of the random 
ensemble of trajectories with different outcomes.

The most basic characterization of the evolving states is via their entanglement entropy.
For free fermions, any non-zero monitoring rate destroys volume-law entanglement \cite{cao2019entanglement},
 in contrast to monitored interacting systems which show  a stable volume law phase  \cite{skinner2019measurement,li2018quantum}.
However, recent literature has shown that
in some cases  free fermion models  
(potentially without number conservation) exhibit non-trivial phase transitions, and phases with  super-area-law entanglement~\cite{nahum2020entanglement,
alberton2021entanglement,jian2022criticality,alberton2021entanglement,turkeshi2021measurementinduced,buchold2021effective,sang2021measurement,kells2021topological,fidkowski2021dynamical,turkeshi2022enhanced,turkeshi2022entanglement,piccitto2022entanglement,muller2022measurement,turkeshi2022entanglementand,lucas2022generalized,coppola2022growth,gal2022volume,merritt2022entanglement,chen2020emergent,kells2021topological,paviglianiti2023multipartite,turkeshi2024density,leung2023theory,ravindranath2023free, piccitto2023entanglement,russomanno2023entanglement,piccitto2024impact}.
Even when the system is in an area-law phase, there may be interesting universal crossovers at intermediate scales.

Most previous work has focused on providing numerical evidence supporting the presence of phase transitions, while analytic results were restricted to certain special Majorana circuits~\cite{nahum2020entanglement,merritt2022entanglement,sang2021entanglement} or simplified settings where the randomness of quantum measurements is eliminated by complete postselection of the outcomes~\cite{turkeshi2021measurementinduced, turkeshi2022entanglementand,gal2022volume,granet2022volume}. 
Very recently, however, analytic insight  has been obtained 
using field-theoretic descriptions for monitored free fermions~\cite{fava2023nonlinear,jian2023measurement,poboiko2023theory}, see also~\cite{yang2023keldysh, poboiko2024measurement, chahine2023entanglement, starchl2024generalized}. 
The starting point of these works is  to use the replica trick to average over the randomness, 
leading to an effective dynamics for $N$ identical copies (or replicas) of the system. 
In analogy to the construction for monitored random circuits~\cite{jian2020measurement,bao2020theory,nahum2021measurement},
which builds on replica approaches to unitary circuits~\cite{zhou2019emergent} and tensor networks~\cite{vasseur2019entanglement}, and related mappings~\cite{hayden2016holographic,nahum2018operator},
the replica trick introduces a symmetry. 
For free fermions, the symmetry group  allows for continuous rotations between replicas~\cite{nahum2021measurement,bao2021symmetry,jian2022criticality,buchold2021effective}.
The continuous replica symmetries give a simple way of understanding the very different phenomenology as compared to interacting systems \cite{nahum2021measurement}, and in some cases the corresponding field theories allow controlled renormalization group approaches.

In this work we derive the replica continuum theory for fermions with number conservation [U(1) conservation] using the approach developed in Ref.~\onlinecite{fava2023nonlinear}
in the setting of Majorana chains.
When applied to the U(1)--conserving case, this gives a controlled derivation of a field theory that  includes an ``entanglement sector''  
(a replica nonlinear sigma model, or NL$\sigma$M) that is coupled to other fields describing the hydrodynamics of the conserved density. 

This derivation clarifies some vexed points in the recent literature.
The fate of U(1)-conserving chains was initially the subject of controversy (about an area law versus a critical state)~\cite{cao2019entanglement,alberton2021entanglement,buchold2021effective, muller2022measurement,coppola2022growth}. Progress was made recently in
Ref.~\onlinecite{poboiko2023theory}, where it was appreciated that the ultimate fate of such chains is in fact an area law state, 
but with an exponentially large correlation length when the monitoring is weak. The SU($N$) NL$\sigma$M that we derive for the entanglement sector was proposed previously in 
Ref.~\onlinecite{poboiko2023theory},
using a different microscopic model:
 while we agree with the qualitative conclusions of that reference, 
we  clarify when this NL$\sigma$M is appropriate. We find that the specific model studied in Ref.~\onlinecite{poboiko2023theory} (and in Refs.~\cite{poboiko2024measurement, chahine2023entanglement}) in fact has a larger replica symmetry, leading to a different NL$\sigma$M. (See also the related discussion of symmetries in random tensor networks in Ref.~\cite{jian2022criticality}.)
We also analyze in more detail the effect of the  conventional hydrodynamic charge mode on the entanglement sector.

In  Ref.~\onlinecite{fava2023nonlinear} the replica trick is used to map the dynamics of a model with random-in-time hoppings
onto the imaginary time evolution of an effective spin chain. 
The use of the number of flavors as a large parameter
then allows us to perform a controlled continuum limit.
For generic Majorana models, i.e. models without U(1) conservation, this led to a nonlinear sigma model (NL$\sigma$M) for an orthogonal $N\times N$ matrix, in the replica limit ${N\to 1}$ \cite{fava2023nonlinear}. 
(The same field theory was also proposed on  grounds of replica symmetry in Ref.~\cite{jian2023measurement}.)
Analysis of this continuum theory then gave an analytical demonstration of the existence of a stable nontrivial phase with  $(\log L)^2$-scaling of the bipartite entanglement ($L$ being the system size), 
separated from the area law phase by a transition that may be driven by dimerizing the measurement rates.

These conclusions of Ref.~\onlinecite{fava2023nonlinear}
crucially depend on the symmetries (or lack thereof) of the initial Hamiltonian and the measurement process. This is in analogy to 
Anderson localization, where there is a well-known symmetry classification \cite{evers2008anderson},
which has recently been extended to Gaussian tensor networks \cite{jian2022criticality}.

For a  U(1)--conserving model with density measurements,
and no further symmetry constraints,
we obtain a continuum Lagrangian of the form (similar to that discussed in Ref.~\onlinecite{poboiko2023theory})
\be
\mathcal{L} = \mathcal{L}_{{\rm NL}\sigma{\rm M}}[Q, {\bf n}]
+ 
 \mathcal{L}_{\rm ferro}[{\bf n}].
\ee
Here $\mathcal{L}_{{\rm NL}\sigma{\rm M}}$ is a NL$\sigma$M for an SU($N$) matrix  $Q(x,t)$, as opposed to an SO($N$) matrix in the Majorana case described above.
The field ${\bf n}(x,t)$, 
which formally resembles the order parameter ${\vec{n} = (n_x, n_y, n_z)}$ of an SO($3$) Heisenberg ferromagnet,
encodes hydrodynamics of the charge (including fluctuations of the charge between different  trajectories).
The coupling between the two sectors is via the coupling constants of the NL$\sigma$M, which depend on the local charge density.

If the hopping amplitudes of the Hamiltonian obey a certain reality condition, then the replica symmetry is enhanced, and a different NL$\sigma$M applies. 
A gauge transformation shows that the reality condition is obeyed for any 1D nearest-neighbor chain with time-independent hoppings,
or for any bipartite model with real hoppings. This includes the models of Refs.~\cite{poboiko2023theory,poboiko2024measurement, chahine2023entanglement}.
We give a more heuristic discussion of this case which suggests that the appropriate NL$\sigma$M is that on the manifold $\mathrm{SU}(2N)/\mathrm{Sp}(2N)$ with ${N\to 1}$.
(The corresponding ${N\to 0}$ theory is also discussed in Ref.~\cite{jian2022criticality}.)
It is worth noting that this does not qualitatively alter the physical conclusions of Ref.~\onlinecite{poboiko2023theory}, since the $\mathrm{SU}(2N)/\mathrm{Sp}(2N)$ NL$\sigma$M  and the $\mathrm{SU}(N)$ NL$\sigma$M have similar properties when $N\to 1$.

Contrary to the Majorana case, the SU($N$) NL$\sigma$M flows under RG to strong coupling, regardless of the monitoring rate.
Therefore  the NL$\sigma$M features no critical phase and, therefore, no phase transition. 
[The same conclusions holds when one consider the group ${\rm SU}(2N)/{\rm SO}(2N)$ instead of SU($N$) for the  NL$\sigma$M.]
Consistently, we predict that the bipartite entanglement entropy satisfies an area-law for all monitoring rates (in one dimension).

Nonetheless, the correlation length is exponentially large at small monitoring rate \cite{poboiko2023theory}
or for a model with a large number of flavors.
When the system size is much smaller than the correlation length, 
the scaling of entanglement is nontrivial: following Ref.~\cite{fava2023nonlinear} we show how scaling forms for the bipartite entropy and for the purity can be obtained using ``RG-improved'' semiclassical computations of the path integral.
In this regime we highlight that there can be a nontrivial effect of charge diffusion on the entanglement.

To end this introduction, we comment on motivations for studying these free fermion problems.

The measurement-induced phase transition in interacting systems can be motivated by practical considerations of computational complexity \cite{skinner2019measurement,napp2022efficient,suzuki2023quantum}, as a transition between trajectories that are classically simulable with polynomial resources and trajectories that require exponential resources.
 In contrast, free fermions can be simulated with polynomial resources regardless of the entanglement structure (though finer distinctions between polynomial scaling laws could be made).
 
Incidentally, this polynomial scaling for free fermions means that large-scale experiments are possible in principle.
Experimental study of quantum trajectories 
\cite{noel2022measurement,koh2023measurement} requires either  ``postselection'' that imposes a prohibitive sampling cost at large size, or auxiliary classical simulation \cite{barratt2022transitions,lee2022decoding,li2023decodable,garratt2023measurements,li2023cross,feng2023measurement}: this auxiliary simulation is feasible in the free fermion case.

Nevertheless, our principal motivation for studying monitored free fermions is theoretical rather than practical.
Free fermions form an important corner of theory space from which we can try to venture out into interacting systems. 
Strikingly, free fermions also provide an elegant physical interpretation for a class of quantum field theories that did not previously have one, in particular a wide range of nonlinear sigma models in the ${N\to 1}$ limit).
The analogies with Anderson localization that exist in the free case may lead to insights into monitored systems (or perhaps even the reverse).
Finally, free systems may be a tractable starting point for obtaining mathematically rigorous results for monitored systems, which are currently lacking.

\tableofcontents

\section{Microscopic model}
\label{sec:microscopic-model}

We will construct a free-fermion model with no global symmetries except charge conservation, 
and with a single dimensionless control parameter:  the  ratio $\Gamma/\mJ$
of the monitoring rate $\Gamma$ and the hopping strength $\mJ$.
To make the model analytically tractable,
we take the couplings in the Hamiltonian to fluctuate like white noise;  we consider the limit of a continuous-time measurement process; and we consider the limit of a large number $N_F$ of fermion ``flavors'', which justifies a semiclassical derivation of the effective theory for any value of $\Gamma/\mJ$.

None of these choices affects the universal features of the results,
which apply to a larger family of models with the same symmetry, including
 the  single-flavor limit ($N_F\to 1$) of the model defined below, which takes the schematic form
\be\label{eq:Hphysnoflavor}
 H(t) = \sum_{j} \lf 
  J_j(t) c_{j}^\dag c_{j+1} 
 + {\rm h.c.} 
 \ri,
\ee
with complex hoppings, together with continuous-time measurements of the local densities $c_j^\dag c_j$ at a rate $\Gamma$. 
In fact,
if $\Gamma/\mJ$ is small, then our derivation  captures not only the universal physics of this single-flavor model but also the quantitative values of the couplings in the effective theory.

An important caveat is that if the Hamiltonian is fine-tuned, the effective description can be modified due to an enlargement of the replica symmetry.
Surprisingly, this occurs for the 1D chain with nearest-neighbor hoppings $J_j$ that are \textit{constant} in time, regardless of the phases of these hoppings.
We will discuss this alternative symmetry class in Sec.~\ref{sec:enlargedsymm} after discussing the more generic case represented by Eq.~\ref{eq:Hphysnoflavor} and its ${N_F>1}$ generalization below.

We consider a  chain of sites ${j=1,\ldots L_{\rm sys}}$, each hosting $N_F$ complex fermion orbitals, with  creation operators 
$c_{j,\mu}^\dag$ labelled by the flavor index ${\mu=1,\ldots N_F}$. The unitary part of the  dynamics is governed by the Hamiltonian
\be
\label{eq:Hphys}
 H(t) = \sum_{j,\mu,\nu} \lf 
  J_j^{\mu\nu}(t) c_{j,\mu}^\dag c_{j+1,\nu} 
 + {\rm h.c.} 
 +   h_{j}^{\mu\nu}(t) c_{j,\mu}^\dag c_{j,\nu} 
 \ri.
\ee
The nearest-neighbor hopping matrix elements $J_j^{\mu\nu}$ and onsite matrix elements ${h_j^{\mu\nu}}$
are complex (with $h_j^{\mu\nu}=\overline{h_j^{\nu\mu}}$) and will be taken to be  white noises.
Initially the unitary dynamics is characterized by two parameters: a hopping rate  $\mJ$,
which sets the white-noise strength  for the inter-site hoppings $J_j^{\mu\nu}$,
and another white-noise strength for the on-site matrix elements $h_j^{\mu\nu}$.\footnote{The mean values of $J$ and $h$ are zero. The nonzero variances are ${\mathbb{E} \left[ |\dd J^{\mu\nu}_j|^2 \right]=\mJ \dd t}/{N_F}$ and 
${\mathbb{E} \left[ |\dd h^{\mu\nu}_j|^2 \right] =\mathfrak{h} \dd t/{N_F}}$, 
where e.g. $\dd J = \int_t^{t+\dd t} J(t')\dd t'$. Note that $\mJ$ and $\mathfrak{h}$ have the dimensions of frequency, i.e. they can be viewed as  rates.}
However we will soon take the limit where the variance for the on-site matrix elements $h_j^{\mu\nu}$ tends to infinity,
so that the unitary dynamics is characterized only by the typical hopping rate~$\mJ$.

We note that the onsite matrix elements $h_j^{\mu\nu}$ only have a nontrivial effect for ${N_F>1}$: 
for ${N_F=1}$ these terms commute with the measured operators, and as a result they can be gauged away by redefining the inter-site hoppings. This is why they were not written in Eq.~\ref{eq:Hphysnoflavor}.

Monitoring modifies this unitary dynamics. 
The monitored observables are taken to be the local densities $\hat n_j^\mu=c_{j\mu}^\dag c_{j\mu}$.
A ``quantum trajectory'' is labelled by the time-series $M_{j}^\mu(t)$ of random measurement outcomes  for each density. 
Given the hoppings and measurement outcomes, the evolution of the state is 
\be
\rho(t) = \f{K(t) \rho_0 K^\dag(t)}{\Tr \left[ K(t) \rho_0 K^\dag(t) \right]},
\ee
where the Kraus operator ${K(t) = \mathcal{T} e^{-i \int_0^{t} \dd t'\, H_{\rm meas}(t')}}$ is the time-ordered exponential of a \textit{non}-Hermitian ``Hamiltonian'':\footnote{{The shift of the number operators by $-{1}/{2}$ and the constant ${-i{\Gamma L_{\rm sys} N_F}/{4}}$ ensure that the Kraus operator are properly normalized: $\mathbb{E}_G[K^\dag(t)K(t)]= \mathbb{I}$.}}
\be
\label{eq:non-hermitian-H-1-replica}
H_{\rm meas}(t)  = H(t) + i \sum_{j\mu} M_j^\mu(t)  
\Big( c_{j\mu}^\dag c_{j\mu}- \frac{1}{2}\Big)
 -i \f{\Gamma N_F L_{\rm sys} }{4}
\ee
Given a realization of the couplings, the probabilities for the measurement outcomes $M$ are given by Born's rule.
The physical Born-rule averages $\mathbb{E}[\bullet]$
may be expressed in terms of simpler Gaussian averages~${\mathbb{E}_{\rm G}[\bullet]}$: 
\be\label{eq:relationbetweenaverages}
\mathbb{E}[\bullet] = \mathbb{E}_{\rm G}\left[ \, \bullet  \, \Tr\lf K(t) \rho_0 K^\dag(t)\ri \right].
\ee
The quantity $\bullet$ may depend on the full  trajectory up to time $t$, but we will usually take it to be a function of the state at time $t$.
In the Gaussian average, the complex variables $J$ and $h$ and the real variable $M$ are all treated as white noise, with respective variances proportional to\footnote{See the previous footnote for variances of $J$ and $h$. For $M$ we take ${\mathbb{E}_G \left[ (\dd M_j^\mu)^2 \right]={  \Gamma  }\dd t}$.
} $\mJ$, $\mathfrak{h}$ (which we take to infinity) and $\Gamma$. Here $\Gamma$ is the measurement rate.
For more details on the above formalism see Ref.~\cite{fava2023nonlinear}; an alternative formulation is via the  stochastic Schr\"odinger equation (see App~\ref{app:stochastic-schroedinger-formulation}).

\section{Effective spin chain}
\label{sec:effective-spin-chain}

In the present limit of large variance for the onsite terms $h_j^{\mu\nu}$, we effectively perform  Haar-random onsite $\mathrm{U}(N_F)$ rotations in every infinitesimal time-step.
This on-site randomization  simplifies the replica description  below.
For a pedagogical introduction to the replica formalism in a similar setting, see  Ref.~\onlinecite{fava2023nonlinear}: 
here we give a very brief outline.

For notational convenience we treat the initial density matrix $\rho_0$  as a \textit{wavefunction} for a ``doubled'' system,  denoting it $\ket{\rho_0}$.
We define an averaged evolution for 
the $N$th tensor power of this density matrix:\footnote{Eq.~\ref{eq:replicatedrhoevolution} assumes a choice of local basis, in order to define the complex conjugation of $K$ and in order to map density matrices to states, but the final results are independent of this choice. We discuss the replicated Hilbert space more carefully in App.~\ref{app:sec:fermionized-replicas}.}
\ba\label{eq:replicatedrhoevolution}
\mathbb{E}_{\rm G} \bigg[ (K(t) \otimes K(t)^*)^{\otimes N} \bigg] 
\ket{\rho_0^{\otimes N}}
& = e^{-  t N_F \mathcal{H}} \ket{\rho_0^{\otimes N}}.
\end{align}
The average on the left is the simple Gaussian average.
On the right, we have taken the average explicitly to give an effective evolution, which involves a time-independent effective Hamiltonian $\mathcal{H}$
 that we specify below.\footnote{In the stochastic Schr\"odinger picture, these effective Hamiltonians are the association $N$ replica Lindbladians (up to an additive constant).}
(The constant $N_F$ has been extracted for convenience.) Note that the effective evolution resembles what would usually be called ``imaginary time'' evolution, although the physical evolution we consider is in real time.

Eq.~\ref{eq:replicatedrhoevolution} allows
physical averages at time  $t$ (taking account of Born's rule,   Eq.~\ref{eq:relationbetweenaverages}) to be written
as analytic continuations $N\to 1$ of transition amplitudes ${\bra{\mathcal{C}} e^{-t N_F \mathcal{H}}\ket{\rho_0^{\otimes N}}}$. The required final state $\bra{\mathcal{C}}$ depends on the quantity of interest. Ultimately these amplitudes may be written as path integrals for fields with specified boundary conditions~\cite{fava2023nonlinear}.

Now we consider the effective Hamiltonian $\mathcal{H}$.
Each physical fermion orbital $c_{j\mu}$
gives rise to $2N$ ``replicated'' orbitals $c_{j\mu}^{\alpha}$.
We label these by  ${\alpha=(\sigma,a)}$ with $\sigma=\pm$ and ${a=1,\ldots,N}$,  where  $\sigma$ distinguishes between
replicas associated with $K$ and $K^*$ operators in Eq.~\ref{eq:replicatedrhoevolution}. 
A choice of convention is required, as for example we are free to make a particle-hole transformation on the ${\sigma=-}$ replicas. In our convention (details in App.~\ref{app:sec:fermionized-replicas})
the total number of replicated fermions on a given site, $\aleph_j=\sum_{\alpha,\mu} c^{\alpha\dag}_{j\mu} 
c^\alpha_{j\mu}$, will be a  constant of motion of $\mathcal{H}$, and  fixed to $NN_F$ by the initial condition in Eq.~\ref{eq:replicatedrhoevolution}.
By contrast ${\sum_{\alpha,\mu} \sigma_\alpha c^{\alpha\dag}_{j\mu} 
c^\alpha_{j\mu}}$, which differs by  the sign factor ${\sigma_\alpha}$, is a dynamical degree of freedom that we relate below to  moments of the physical charge density $\hat n_j$.

The effective Hamiltonian $\mathcal{H}$ is  quartic in the fermions as a result of the Gaussian averaging in Eq.~\ref{eq:replicatedrhoevolution}. In the limit of interest where the on-site noise in (\ref{eq:Hphys}) is strong,
$\mathcal{H}$ may be written in terms of bosonic ``spin'' operators 
\be
S^{\alpha\beta}_j = 
\f{1}{N_F} 
\lf
\sum_\mu
c_{j\mu}^{\alpha\dag}c_{j\mu}^{\beta} - \f{\delta^{\alpha\beta}}{2N}
\aleph_j
\ri.
\ee
These operators  
are $\mathrm{su}(2N)$ generators, normalized so that ${
[{S}_j^{\alpha\alpha'}, {S}_j^{\beta\beta'}] = {N_F}^{-1}( \delta_{\alpha',\beta} {S}_j^{\alpha\beta'} - \delta_{\alpha\beta'} {S}_j^{\beta\alpha'} )}$.
Defining ${\hat \Gamma :=\Gamma N_F/(N_F+1)}$, the effective Hamiltonian is (repeated indices summed):
\ba
\label{eq:Hamiltonian-final}
    \mathcal{H} &= \sum_j \left[ -\f{\mathcal J}{2} {S}_j^{\alpha \beta} {S}_{j+1}^{\beta\alpha} 
    + \f{\hat \Gamma \sigma_\alpha \sigma_\beta }{2} 
    \left({S}^{\alpha \beta}_j {S}^{\beta\alpha}_j - {S}^{\alpha\alpha}_j {S}^{\beta\beta}_j \right) \right].
\end{align}
The strong on-site noise has projected the dynamics into a subspace of the original replicated Hilbert space:
each site hosts a $\mathrm{su}(2N)$ spin in a definite representation,  corresponding to a rectangular Young tableau with $N$ rows and $N_F$ columns.\footnote{These states are invariant under the action of the $\mathrm{su}(N_F)$ flavor (as opposed to replica) symmetry. The $su(N_F)$ generators are $T_{\mu\nu}=\sum_\alpha c_\mu^{\dag\alpha}c_\nu^\alpha -\frac{\delta_{\mu\nu}}{N_F}\aleph$. They  commute with the $\mathrm{su}(2N)$ generators.
With this representation,  large $N_F$ gives a semiclassical limit \cite{affleck1985quantum} 
} A derivation of $\mathcal{H}$ is given in App.~\ref{app:derivation-Hamiltonian}, including an additive constant that is omitted in 
Eq.~\ref{eq:Hamiltonian-final}
and which ensures that the ground state energy of $\mathcal{H}$ is exactly zero
when ${N=1}$, as required for the generator of a quantum channel.

For ${N=1}$, Eq.~\ref{eq:Hamiltonian-final} retains full ${\mathrm{su}(2N)=\mathrm{su}(2)}$ symmetry  for any measurement rate,\footnote{Because the $\hat \Gamma$ term reduces to a constant at $N=1$.} 
but for general $N$ and $\hat \Gamma$  it  preserves only ${\mathrm{su}(N)\oplus \mathrm{su}(N)\oplus u(1)}$, as will be clear in sub-block representation below.

\subsection{Semiclassical ground states}

While they remain finite in the limit ${N_F\rightarrow \infty}$, the operators $S_j^{\alpha\beta}$ commute in this limit. 
The choice of  representation also imposes kinematic constraints on these matrices of the form\footnote{A simple check on this constraint consists in verifying it when acting on the on-site identity state, i.e. checking $S_j^{\alpha\beta} S_j^{\beta\gamma} |\mathbb{I}\rangle = \f{\delta_{\alpha\gamma}}{4}|\mathbb{I}\rangle + O(N_F^{-1})$, and then arguing that any 
state
with $\aleph_j=NN_F$ is in the $\mathrm{SU}(2N)$--orbit of 
$|\mathbb{I}\rangle$. Here $\ket{\mathbb{I}}$ is (up to a choice of normalization) the replicated infinite-temperature density matrix, represented as a ket in the way discussed around Eq~\ref{eq:replicatedrhoevolution}: see App.~\ref{sec:boundar_states}.} 
 ${S_j^{\alpha\beta} S_j^{\beta\gamma} = \f{\delta_{\alpha\gamma}}{4} \mathds{1} + O(1/N_F)}$,
which specify the semi-classical phase space.

Given that the average over noise and measurement outcome turns the time-evolution problem into the imaginary time evolution generated by $\mathcal{H}$, we are interested in understanding the nature of the ground state of $\mathcal{H}$. For this purpose, we consider first the classical ground states and will later characterize quantum fluctuations on top of them.

At the classical level $S_j$ is a Hermitian matrix which we may decompose as 
\ba \label{eq:semiclassical-variables}
    S_j &=
    \begin{pmatrix}
        {L}_j+ d_j \mathds{1}  & q_j{Q}_j\\
        {Q}_j^\dag q_j^* & {R}_j - d_j \mathds{1}
    \end{pmatrix}.
\end{align}
These variables satisfy ${\tr L_j=\tr R_j =0}$ and ${\det Q_j=1}$,\footnote{The decomposition is not fully unique as for ${N>1}$ we can multiply $q^{-1}$ and $Q$ by an $N$th root of unity, but this will not affect the perturbative analysis of the  sigma model below.}
as well as the kinematic constraint above, which reads 
\ba\label{eq:LQRconstraints}
    L_j Q_j + Q_j R_j&=0,\\
    (L_j+d_j\mathds{1})^2 + |q_j|^2 Q_j Q_j^\dag &= \mathds{1}/4,
\end{align}
up to $O(1/N_F)$ corrections.

The variable $d_j$ is related to the physical charge ${n_j=N_F^{-1}\sum_\mu c_{j\mu}^\dag c_{j\mu}}$ in the unreplicated theory. This is best seen in terms of the quantum operator $\hat{d}_j = \sum_\alpha \sigma_\alpha S_j^{\alpha\alpha}/(2N)$ corresponding to the classical variable $d_j$, which satisfies
\be\label{eq:dchargerelation}
    \lim_{N\to 1}\braket{\mathbb{I}| \hat{d}_j^k |\rho^{\otimes N}} = \Tr\Big[ \rho \Big( n_j-\f{1}{2} \Big)^k \Big].
\ee
We will refer to the variables $(d,q)$ as the charge sector and $(Q,L,R)$ as the entanglement sector.

Inserting the block decomposition, the Hamiltonian~\eqref{eq:Hamiltonian-final} takes the simple form (up to a constant)
\ba
\label{eq:spin-Hamiltonian-explicit}
    \mathcal{H} = 
    \sum_j\! \Big[
     - \f{\mJ}{2}\! \tr\left(S_j S_{j+1}\right) + \hat{\Gamma}  \tr\left( L_j^2 + R_j^2\right) 
    + 2 N\hat{\Gamma}_N d_j^2 \Big].
\end{align}
with $\hat{\Gamma}_N=\hat{\Gamma}(1-N)$. Due to the first --- ferromagnetic --- term, the classical ground states are translation-invariant and, because of the second term, they satisfy ${L=R=0}$. This gives the manifold 
\ba
\label{eq:manifold}
    Q^\dag Q  & = \mathds{1},
    &
      L&=R=0,
      &
    d^2 + |q|^2 = {1}/{4},
\end{align}
i.e. $Q\in \mathrm{SU}(N)$.
The variables $(d,q)$ in the charge sector parameterize a sphere: in the limit  ${N\to 1}$, all points on this sphere become degenerate. 
We will also parameterize this sphere using a unit vector ${{\mathbf n} = (q+q^*, i(q^*-q), 2d)}$.
In a translationally-invariant state the value of $d$ is fixed by the charge density, $d=\sum_j \lf n_j-1/2\ri /L_{\rm sys}$.

\section{Mapping to $\mathrm{SU}(N)$  sigma model}
\label{sec:NLSMmapping}

The variables $Q$, $d$, and $q$, which parameterize the classical ground states, are the slow modes. 
Starting from a coherent state path-integral and integrating out the gapped modes $L$ and $R$, one could obtain the effective Lagrangian describing the large-scale fluctuations of $(d,q,Q)$.
As a shortcut, we shall infer the effective Lagrangian from the semiclassical equations of motion (as in  Ref.~\onlinecite{fava2023nonlinear}). 
The derivation is quantitatively controlled in the semi-classical limit $N_F\gg 1$, but we expect that the universal consequences hold for all $N_F$, as they are largely determined by the symmetries of the effective model.

The semiclassical equations are obtained as the classical limit of the Heisenberg equations in App.~\ref{app:Heisenberg-equations}.
Because of the large parameter $N_F$,
the continuum limit can be taken in a controlled way.\footnote{The parameter $N_F$ stands outside the action in a coherent states path integral. As a result, fluctuations of ``massive'' modes $L$ and $R$
are small ($L^2\sim R^2\sim N_F^{-1}$). Similarly, large-momentum fluctuations of the ``Goldstone'' modes are strongly suppressed.}
This amounts to expanding the equations of motion both in spatial derivatives and in powers of the ``massive'' modes $L$ and $R$: 
details are given in App.~\ref{app:Heisenberg-equations}. For convenience we make a Wick rotation during this intermediate step, $t=i\tilde t$, so that the time evolution resembles Schrodinger evolution in the $\tilde t$ coordinate; below, $\dot Q = \partial_{\tilde t} Q$ etc.

The Heisenberg equations are first order in time derivatives: that for $Q$ is  of the form\footnote{Above, we  neglect the $O(\partial_x^2)$ term because $\partial_t Q$ is of the same order as $\partial_x Q$, as will be clear below; see App.~\ref{app:Heisenberg-equations} for a more careful derivation.}
${\dot Q = 4i\hat \Gamma L Q + O(\partial_x^2)}$
together with an expression for $\dot L$.
Eliminating $L$ by taking another time derivative, the equation of motion for the entanglement sector is
\be
\label{eqs:entanglement-eom}
Q^\dag \left[
\ddot Q - 4 \mathcal{J}\hat \Gamma \, 
\partial_x  \big( |q|^2 \partial_x Q \big) 
\right]  - \rm{h.c.} = 0 ~.
\ee
The equations of motion for the charge sector are
\ba\label{eqs:charge-eom}
    {i} \dot{q} & =  \!  \mJ \! \left(  q\partial_x^2 d -  d \partial_x^2 q  \right) 
    \!+\! q d \left[ 
    \f{\mJ }{N}\!  \tr { \partial_x Q^\dag \partial_x Q }           
    - 4\hat{\Gamma}_N 
    \right]\!, \notag
    \\
    {i}\dot{d} & = \f{\mJ}{2}\! \left( q^* \partial_x^2 q - q \partial_x^2 q^*   \right)\!,
 \end{align}
plus terms of order $O(\partial_x^{3}, L \partial_x^{2})$ in both equations.

Equations (\ref{eqs:entanglement-eom},\ref{eqs:charge-eom}) are the equations of motion of an action which comprises 
a nonlinear sigma model for the unitary matrix $Q\in \mathrm{SU}(N)$, together with the action for a ferromagnetic spin chain in the charge sector.\footnote{The overall coefficient of $\mathcal{L}$ 
cannot be inferred from the equations of motion, but is fixed by noting that the terms without time derivatives are inherited directly from the Hamiltonian in the coherent states approach \cite{fradkin2013field}.}
Returning to the physical time coordinate $t$,  writing the path integral weight as $\exp(-\int \dd t \dd x \mathcal{L})$, 
and writing $(q,d)$ in terms of the unit vector ${\mathbf n}$,
\begin{align} \label{eq:action-N}
\mathcal{L} &=  \mathcal{L}_{{\rm NL}\sigma{\rm M}}[Q,q] +  \mathcal{L}_{\rm{ferro}}[\vec{n}].
\end{align}
The NL$\sigma$M Lagrangian is given  by 
\ba\label{eq:nlsmmaintext}
\mathcal{L}_{{\rm NL}\sigma{\rm M}}  &=    \f{1}{{2g_B}} \,\tr\!\Big(v^{-1}\partial_t Q^\dag \partial_t Q
    + v\,\partial_x Q^\dag \partial_x Q\Big),
\end{align}
with a ``bare'' local coupling constant $g_B$ and local velocity $v$ that depend on $|q|^2$:
\ba\label{eq:gandv}
g_B & = \f{2\sqrt{\hat \Gamma}}{N_F |q| \sqrt{\mJ}}, 
&
v  & = 2 |q| \sqrt{\hat \Gamma \mJ}.
\end{align}
Using Eq.~\ref{eq:manifold},
and writing $d$ in terms of the local charge density $n$ via ${d=n-1/2}$ (see eq.\eqref{eq:dchargerelation}),
we may further write 
\be
|q| = \sqrt{n(1-n)}.
\ee
The dependence of Eqs.~\ref{eq:gandv} on $|q|^2$ constitutes the  interaction between the charge and entanglement sectors.
The Lagrangian for the ferromagnet in \eqref{eq:action-N} is 
\be
\label{eq:L-ferro}
    \mathcal{L}_{\rm{ferro}} = 
    \frac{{NN_F } }{2} \left[
    i (1-n_z) \dot{\varphi} + \f{\mJ}{4} \left| \partial_x \vec{n} \right|^2 + \hat{\Gamma}_N n_z^2 \right].
\ee
The first term is the Berry phase for a chain of $\rm{su}(2)$ spins of size ${N N_F/2}$ and unit lattice spacing, expressed in terms of $\varphi$ --- the angle between $(n_x,n_y)$ with the $x$-axis. 
Note that ${\lim_{N\to 1}\hat \Gamma_N=0}$,
so that $\rm{su}(2)$ symmetry of the charge sector is recovered
in the replica limit.
 This matches the well-known formulation of charge transport in the classical symmetric exclusion process via the $\rm{su}(2)$ ferromagnet \cite{thomas1980quantum,schutz1994non,tailleur2008mapping,10.21468/SciPostPhys.12.1.042}. 

Above we obtained precise quantitative results for the couplings
by taking ${N_F\gg 1}$.
In fact, we expect that  these results remain quantitatively accurate (at sufficiently large scales) even when ${N_F=1}$, if we impose the additional condition ${\Gamma\ll\mJ}$.
This is because 
in the unitary limit ${\Gamma=0}$
the spin chain Hamiltonian (\ref{eq:Hamiltonian-final}) becomes an $\mathrm{su}(2N)$ symmetric ferromagnet, whose ground states are simple polarized product states. 
For small $\Gamma$, the low-energy-density states remain locally almost-polarized, so  large polarized blocks act like effective spins with a large effective~$N_F$.\footnote{\label{footnote:ferrocrossover} Recall that in the usual $\mathrm{su}(2)$ ferromagnet, semiclassics is accurate at low energy densities even in the absence of any other small parameter \cite{takahashi1987few}. The heuristic reason for this is quoted above. Formally, we can make a coarse-graining argument in which we obtain a large parameter $b$ outside the coherent states action by rescaling lengths by a factor of $b$ (and times by a factor $b^2$).
If  ${\Gamma\ll \mJ}$, the same argument shows that the present spin chain   resembles a pure ferromagnet on lengthscales smaller than $b_*\sim \sqrt{\mJ/\Gamma}$
(at which point the $\Gamma$ term becomes comparable with the other terms in the action). By coarse-graining to the scale $b_*$, we acquire an effective ``$N_F$'' given by $b_*$. This justifies the derivation given in the text, even if microscopically $N_F$ is equal to 1. (Note that, although in this discussion $\Gamma$ is assumed to be small, the fact that it is nonzero is important, since on scales $\gtrsim b_*$ it gives a mass to the $L$ and $R$ modes and leads to the effective description in the text.)}

If we impose neither ${N_F\gg 1}$ nor ${\Gamma\ll \mJ}$ then the quantitative (bare) values of $g$ and $v$ cannot be assumed to be accurate, but nevertheless the universal consequences of the above theory, discussed below, remain valid.
On grounds of replica symmetry, the effective description obtained above applies for ``generic'' monitored free fermion models with particle number conservation.

The ${\mathrm{SU}(N\to 1)}$ NL$\sigma$M
in the limit $N\to 1$ was obtained for a microscopic model in a different regime in Ref.~\cite{poboiko2023theory}.
While we agree that this is the correct replica symmetry class for generic conserved fermion models, in fact we find that it does not apply to the specific model studied in Ref.~\cite{poboiko2023theory}, as a result of the hopping amplitudes in that model being real.
Imposing constraints on the phase of the hoppings can lead to an enlarged ${\mathrm{su}(2N)}$ replica symmetry which we discuss in Sec.~\ref{sec:enlargedsymm}. Related symmetry observations were made for the $N\to 0$ limit in Ref.~\cite{jian2022criticality}.

\section{Consequences of the field theory}
\label{sec:consequencesoverview}
 
We start with a schematic overview of the consequences of the continuum description in Eqs.~\ref{eq:nlsmmaintext},~\ref{eq:L-ferro}, with some details to follow.

\subsection{Regimes of scales and RG flows}
\label{sec:regimes}

The first point to note is that the 
nonlinear sigma model coupling $g$ is small if either $N_F\gg 1$ or $\Gamma/J\ll 1$ holds.
This guarantees that there is a large regime of length and time scales where the theory can be treated ``semiclassically'' 
(more precisely, by an RG-improved semiclassics discussed below and in Ref.~\cite{fava2023nonlinear}).

In this ``semiclassical'' regime, various von Neumann entropies/entanglement entropies can be computed essentially by solving the classical equations of motion.
The simplest setting is where the charge density is homogeneous. 
In this case it is possible to argue that we may deal only with  translationally invariant equations for the $Q$ sector.
More generally, we must solve the saddle point equations both for the charge and for the entanglement. 
However, this simplifies considerably for the calculation of the von Neumann entropy:
because the matrix $Q$ becomes trivial in the limit ${N\to 1}$, the entanglement sector does not feed back on the charge sector, and the latter reduces to standard classical hydrodynamics (App.~\ref{app:semiclassicaleqmot}).

Although entropies scale nontrivially in the semiclassical regime, this is an intermediate lengthscale effect:
the dynamics is ultimately in the \textit{disentangling} phase, because --- as discussed in Ref.~\cite{poboiko2023theory} --- the NL$\sigma$M flows to strong coupling \cite{HIKAMI1981208,evers2008anderson}: 
\be\label{eq:betafunction}
\f{\dd g}{\dd  \tau} = \f{N}{4 \pi} g^2 + O(g^3).
\ee
Here $\tau$ is the logarithm of the observation lengthscale.
Below we will denote lengthscales by $L$: this should not be  confused with the operators $L_j$ discussed above, which will not appear again below.
The coupling to the charge fluctuations is irrelevant for this RG flow: 
see the end of Sec.~\ref{sec:computingentropies}.
Integrating this equation for ${N=1}$,
\be\label{eq:runningcoupling}
\f{1}{g(\ln L)} = \f{1}{g_B} - \f{\ln L}{4\pi} ,
\ee
shows that the running coupling $g=g(\ln L)$ becomes of order $1$ at a lengthscale 
\be
\xi \asymp \exp \lf \f{4\pi}{g_B} \ri,
\ee
where $\asymp $ denotes asymptotic equivalence of the logarithms of the two sides. The standard assumption is that on scales larger than $\xi$ the NL$\sigma$M flows to a massive phase \cite{poboiko2023theory}. 
(This is consistent with numerics in  Sec.~\ref{sec:numerics}.)
As a result, the pure states produced by the dynamics at asymptotically late times are \textit{area law} states \cite{cao2019entanglement}, 
for any value of $\Gamma/\mJ$,
though with a correlation length $\xi$ that is extremely large if the bare value of $g$ is small.
We note that rigorous results are lacking in this subject, and it  would be valuable to have a rigorous proof of the existence of a stable area law phase, for example by working with the lattice model in the opposite regime of ${\Gamma\gg \mJ}$,~${N_F=1}$.

The various regimes are illustrated by the form of the bipartite von Neumann entanglement entropy, 
for a pure state at asymptotically late times in the dynamics. 
We consider a system of size $L$, with non-periodic boundary conditions, cut into two halves (Sec.~\ref{sec:computingentropies}).
When $L$ is large compared to microscopic scales\footnote{\label{footnote:microscopiclengthscale} If $\Gamma/\mJ$ is of order 1 then it is sufficient to assume $L\gg 1$ (in order for the long-wavelength theory to be valid).
In the weak-measurement limit  $\Gamma/\mJ\ll 1$
then we must assume  ${L\gg \sqrt{\mJ/\Gamma}}$, as a result of the crossover from the unitary theory discussed in Footnote~\ref{footnote:ferrocrossover} (similarly, for Eq.~\ref{eq:entropydensity} we require $vt\gg \sqrt{\mJ/\hat \Gamma}$).
Ref.~\cite{poboiko2023theory} considers fermions that are ballistic at small scales, also leading to a crossover phenomenon.} but still much smaller than $\xi$, we find
\ba\label{eq:bipartiteentropy}
 S_{\rm vN}(L) &\simeq 
 \f{\pi \log L }{3 g_B} - \f{(\log L)^2}{24} 
&
& (L\ll \xi).
\end{align}
On scales much larger than $\xi$, 
$S_{\rm vN}(L)$ plateaus at an area-law entanglement value of order  $\lim_{L\to \infty} S_{\rm vN}(L) \propto 1/g_B^2$ (which is of order ${g_B^{-1} \ln \xi}$).

Note that Eq.~\ref{eq:bipartiteentropy} takes into account the RG flow of $g$. Neglecting this flow corresponds to 
a straightforward saddle-point calculation using the bare action: this gives the first term, 
$S_{\rm vN}\propto g_B^{-1} \ln L$.
The latter result is valid only when ${[\ln L]/[\ln \xi]\ll 1}$, which is a more restrictive condition than the ${L/\xi\ll 1}$ assumed in Eq.~\ref{eq:bipartiteentropy}.

As another example, the decay of (mixed state) entropy density $s$ \cite{gullans2020dynamical} for a high-entropy initial state (in the thermodynamic limit) takes the form 
\ba\label{eq:entropydensity}
s   & \simeq  \f{\pi^2}{3} \f{1}{g(\log v t)} \f{1}{ v t}
& 
&(t\ll \xi/v).
\end{align}
Again, we have assumed that $t$ is large compared with an appropriate ``microscopic'' timescale (Footnote \ref{footnote:microscopiclengthscale}).

On timescales much larger than $\xi/v$
the sigma model is massive, and the entropy density decays exponentially.

\subsection{Computing entropies}
\label{sec:computingentropies}

The  calculation of entropies  was discussed in Ref.~\cite{fava2023nonlinear} in a closely related setting. This involved finding the minimal action configuration for the boundary conditions corresponding to the system bipartition. Here we mention only the new features of the calculation, referring to Ref.~\cite{fava2023nonlinear} for further detail.

For simplicity let us first consider the regime ${\ln L \ll \ln \xi}$. 
In this regime the running coupling (\ref{eq:runningcoupling})
is approximately equal to the bare coupling: therefore we can neglect the nontrivial RG flow, and do semiclassics using the bare stiffness $1/g_B$.
Let us also first consider the case where the charge density $n$ is uniform (on scales much larger than the lattice spacing).

Computations of entropies (either the mixed state entropy or the entanglement entropy) require us to compute the partition function of the field theory with appropriate boundary conditions for both $Q$ and ${\bf n}$, discussed in App.~\ref{sec:boundar_states} and Ref.~\cite{fava2023nonlinear}.
For concreteness, consider the   mixed state entropy. We must then  compute the  $N\to 1$ limit of a partition function with $Q=\mathds{1}$ at the initial time, and with $Q = { Q_{\rm vN}}$ at the final time where
\be
{\footnotesize { Q_{\rm vN}} \equiv 
\lf \hspace{-2pt} \begin{array}{ccccc}
    \phantom{+} 0 & \phantom{+} 0 & \cdots &  \phantom{+} 0 &  + 1  \\
    -1 & \phantom{+} 0 & \cdots &  \phantom{+} 0 &  \phantom{+} 0  \\
    \phantom{+} 0 & -1 & \cdots &  \phantom{+} 0 & \phantom{+}  0  \\
     &   & \cdots & &     \\
     \phantom{+} 0 & \phantom{+} 0 & \cdots & -1 &\phantom{+}  0
\end{array}\ri. }
\ee
(Up to signs, this matrix represents a cyclic permutation of the $N$ replicas.) In the present setup, the charge sector has a trivial saddle-point solution, in which ${|q|=\sqrt{n(1-n)}}$ is a constant that simply sets $g$ and $v$ in the \NLSM action.
The only nontrivial saddle-point equation is then (\ref{eqs:entanglement-eom}), Wick-rotated back to the physical $t$ coordinate:
\be
\label{eqs:entanglement-eomwick}
Q^\dag \left[
\partial_t^2 Q + 4 \mathcal{J}\hat \Gamma \, 
\partial_x  \big( |q|^2 \partial_x Q \big) 
\right]  - \rm{h.c.} = 0 ~.
\ee
Further, for the present boundary conditions, $Q$ is real.
As a result, the solution lies in the submanifold ${\mathrm{SO}(N)\subset \mathrm{SU}(N)}$, and reduces to that discussed in \cite{fava2023nonlinear}:
$Q=Q(t)$ interpolates smoothly between the initial and final conditions.
Evaluating the action of this saddle point and taking the replica limit (see Sec.~VI of  \cite{fava2023nonlinear}) gives the result in Eq.~\ref{eq:entropydensity}.
Similar considerations give the bipartite entanglement entropy in the present regime  of lengthscales (${\ln L\ll \ln \xi}$).

An additional subtlety in comparison to Ref.~\cite{fava2023nonlinear} is that, for general initial conditions, we must also solve the saddle-point equations for the charge sector \cite{stone2000semiclassical,tailleur2008mapping,unitarymajorana}. 
For general $N$, these are affected by the entanglement sector.
However, we argue that, in the calculation of the von Neumann entropy, 
the replica limit is straightforward for the charge equations of motion. 
We may directly set $N=1$ in these equations, and use the fact that $Q$ becomes a trivial constant in this limit (${Q=1}$). 

We note that this simplification holds for the von Neumann entropy --- for which the boundary conditions have a straightforward ${N\to 1}$ limit --- but may not hold identically for other observables, such as the averaged purity.\footnote{For the boundary conditions appropriate to the von Neumann entropy, 
$Q$ becomes a trivial constant ($Q=1$) in the ${N\to 1}$ limit.
[The quantity $\tr (\partial_x Q^\dag) (\partial_x Q)$ is of order $N-1$ for the solutions we consider (App.~\ref{app:semiclassicaleqmot}).]
For other observables the replica limit may be more complex.
The average purity is an example: in that case, the $N$-dependent boundary condition for $Q$ is a matrix that is defined for $N>1$ but not for ${N=1}$. 
In that case, a priori we should solve the saddle-point equations for both sectors in full for ${N>1}$ and analytically continue only after obtaining the action.
There may be a physical reason for this difference. The average of the purity is the average of $e^{-S_2}$, where $S_2$, the second R\'enyi entropy, is proportional to $1/g_B\asymp N_F$ in the regime we are discussing. Since this large factor appears in the exponent,  inserting the purity into the average may bias the ensemble of quantum trajectories \cite{rakovszky2019sub} away from those that are typical. This effect is probably subleading at large scales, however, since $S_2$ scales at most logarithmically with scale in the present model, while imposing a rare charge fluctuation on a large scale incurs a polynomial cost.}

The charge equations of motion are then those of the (``imaginary time'') ferromagnet \cite{tailleur2008mapping,unitarymajorana}, and are discussed in App.~\ref{app:semiclassicaleqmot}. 
For all the cases that we discuss, they reduce to conventional hydrodynamics of the physical charge density $n$:\footnote{More general final-time boundary conditions (e.g. conditioning on a final-time charge profile as well as an initial-time profile) lead to two nontrivial equations \cite{stone2000semiclassical,tailleur2008mapping, unitarymajorana}, see App.~\ref{app:semiclassicaleqmot}.}
\be\label{eq:chargehydro}
 \partial_t n = \f{\mJ}{2} \partial_x^2 n.
\ee
Above we have restricted to cases where $n$ is simply constant, but  the approach applies more generally.
Note that  charge transport takes place diffusively (dynamical exponent $z_{n}=2$), while the entanglement sector has the dynamical exponent ${z_Q=1}$.

So far we have considered pure semiclassics. 
On larger scales
(even when $L\ll \xi$)
we must take into account the running of the coupling. This is described in Ref.~\cite{fava2023nonlinear}.
For the purification setup (\ref{eq:entropydensity}) it is sufficient to run RG up to the lengthscale $vt$, and then do semiclassics using the renormalized coupling at this scale (\ref{eq:runningcoupling}).
For the bipartite entropy, 
we must integrate over lengthscales at increasing distances from the entanglement cut, 
using the renormalized coupling appropriate to each scale. This integral gives Eq.~\ref{eq:bipartiteentropy}.

Finally, we comment on the role of the charge sector in the renormalization group flow of the \NLSM. We argue that
the coupling between the two sectors is in fact RG-irrelevant,
so that we are justified in using the standard beta function for the NL$\sigma$M (\ref{eq:betafunction}).
Indicating the space-time coordinate by  $\mathbf{y}=(x,t)$ , 
the coupling between the two theories in Eq.~\ref{eq:nlsmmaintext} is of the form $\int \dd \mathbf{y} \, n_z^2(\mathbf{y}) \mathcal{O}(\mathbf{y})$ with $\mathcal{O}=\tr[(\partial_x Q^\dag)\partial_x Q]$. 
We can imagine integrating out $\vec{n}$ to obtain a Lagrangian involving only $Q$.   By the cumulant expansion,
this generates terms of the form 
\be\label{eq:chargecorrections}
\int \dd^2 \mathbf{y}_1 \ldots \dd^2  \mathbf{y}_k\,\Delta^{(k)}(\mathbf{y}_1,\ldots,\mathbf{y}_n)\mathcal{O}(\mathbf{y}_1) \cdots \mathcal{O}(\mathbf{y}_k),
\ee
where $\Delta^{(k)}(\mathbf{y}_1,\ldots,\mathbf{y}_k)$ is the \emph{connected} $k$-point correlation function for $n_z$
 (computed using the Lagrangian of the charge sector)
which decays as a power law at large times/distances. 
(In fact these correlations are Gaussian at leading order and match standard classical fluctuating hydrodynamics for a conserved density.\footnote{ Writing $n_z = \<n_z\> + \delta n_z$, one convenient rewriting of the charge Lagrangian is in terms of $h(x,t) = \int^x_0 \dd x' \delta n_z(x')$. After integrating out the angle in the $(n_x,n_y)$ plane, $h$ has a free field action with dynamical exponent $z_n = 2$.
 As an aside: above we have discussed the case where the charge is at equilibrium. But even if the charge is driven out of equilibrium by connecting to boundary reservoirs at different chemical potentials, 
this leads to a current of order $1/L$ at large $L$, so that in the limit $L\to\infty$ the RG flow for $g$ will be unaffected.})
Since $\mathcal{O}$ is a marginal operator at the $g=0$ fixed point, 
all these terms are \textit{irrelevant}, except for the first one, $\langle n_z^2 \rangle \mathcal{O}(\mathbf{y})$, which simply forms part of the standard translation-invariant NL$\sigma$M action.
Thus we are justified in considering only the RG  for the pure NSLM, in which  $\langle n_z^2\rangle$ ---
which at large $N_F$ is also approximately equal to $\langle n_z\rangle^2$ ---
appears only as a parameter setting the bare couplings.

\subsection{Higher-dimensional models}

The derivation we have given for the continuum field theory extends immediately to higher-dimensional versions of the models in Eqs.~\ref{eq:Hphysnoflavor},~\ref{eq:Hphys}.
In higher dimensions the NL$\sigma$M has a phase transition between a nontrivial phase and an area-law phase \cite{poboiko2024measurement,chahine2023entanglement}. (In $d=1+\epsilon$ spatial dimensions, we can see the presence of a phase transition  at a critical point $g_*$  by including the tree level term, ``$-\epsilon g$'', on the right-hand-side of Eq.~\ref{eq:betafunction}.)
The nontrivial phase at $g<g*$ is the one in which the NL$\sigma$M flows to weak coupling, 
which means that the scaling forms obtained by straightforward semiclassics are valid on asymptotically large lengthscales.  For $g>g_*$ the NL$\sigma$M flows to a massive area-law phase.

In $d$ spatial dimensions this leads to a bipartite entanglement entropy scaling as $\f{1}{g_B} L^{d-1} \ln L$  in the nontrivial phase \cite{poboiko2024measurement,chahine2023entanglement}. 
This scaling 
also holds for a stable phase in a simpler model of projectively measured Majoranas above one dimension \cite{nahum2020entanglement}, for similar reasons
\cite{nahum2021measurement}.

At the higher-dimensional critical point for U(1) fermions, we must check the relevance/irrelevance of  coupling between charge fluctuations and the entanglement sector.
The equivalent issue has been examined recently  for the measurement-induced  critical point in \textit{interacting} quantum circuits \cite{ha2024measurement}, where charge fluctuations are generically RG-\textit{relevant} in 1+1D.

Here, we saw above that charge fluctuations can couple to local $\mathrm{SU}(N)$-invariant operators: in particular, 
the microscopic derivation shows that there is a coupling between the charge density and $\mathcal{O}=\tr[(\nabla Q^\dag)\nabla Q]$.
At the critical point (and at long wavelengths), 
$\mathcal{O}$ is equivalent to the relevant operator which drives the phase transition between trivial and nontrivial phases.
As in the interacting case, the Harris criterion 
shows that the charge-entanglement coupling is irrelevant if ${\nu >2/d}$, where $\nu$ is the correlation length exponent \cite{ha2024measurement}.\footnote{\label{foot:Harris}In the language of Eq.~\ref{eq:chargecorrections}, 
we may obtain this condition by examining the scaling dimension of the term with ${k=2}$, taking into account the scaling dimension ${x=d+1-1/\nu}$ of the relevant operator driving the transition \cite{cardy1996scaling}.
Since the dynamical exponent for the charge is $z=2$ whereas the entanglement critical point has $z=1$ we  approximate the charge as static, i.e. we approximate $\Delta^{(2)}$ as a time-independent spatial delta function.}

The one-loop result in the ${d=1+\epsilon}$ expansion is ${\nu =1/\epsilon}$, so that 
the charge coupling is formally irrelevant for very small epsilon.
It requires simulations to establish what happens for $d=2$ or $d=3$.

In Refs.~\cite{poboiko2024measurement,chahine2023entanglement}, models in ${d=2}$ spatial dimensions were simulated, and 
Ref.~\cite{poboiko2024measurement} found a critical point with   $\nu\approx 1.4$.
In Sec.~\ref{sec:enlargedsymm} we argue that,
because of a symmetry constraint,
the models in Refs.~\cite{poboiko2024measurement,chahine2023entanglement} 
are described by
the sigma model on the target space $\mathrm{SU}(2N)/\mathrm{Sp}(2N)$, 
rather than the sigma model on  $\mathrm{SU}(N)$  as was believed previously.
Therefore these lattice models are slightly different from the models we have been discussing so far in this section.
Nevertheless, we expect a similar effective field theory, coupling charge and entanglement degrees of freedom, in the $\mathrm{SU}(2N)/\mathrm{Sp}(2N)$ symmetry class. 

At first sight, the result 
$\nu\approx 1.4$ then indicates that the charge--entanglement coupling is irrelevant in that symmetry class in $d=2$ spatial dimensions. 
This seems very  plausible, since $\nu$ values larger than 1 are encountered for many related Anderson localization sigma models.
Strictly, however, to draw this conclusion one would need to check that  the critical point found numerically is the one governed by the pure 
$\mathrm{SU}(2N)/\mathrm{Sp}(2N)$ 
model, and not by a different fixed point that is influenced by charge fluctuations.
Evidence that the dynamical exponent was unity would be  an indication that the transition is indeed governed by the pure \NLSM.

The discussion above applies to the critical point. One could further ask if the $\vec{n}$-$Q$ coupling can affect the entanglement scaling in the ordered phase  (the nontrivially entangled phase). Proceeding as in footnote~\ref{foot:Harris}, it is easy to check that the $\vec{n}$-$Q$ coupling is irrelevant in the ordered phase.

\section{Numerical results}
\label{sec:numerics}

In this section, we confirm the drift towards ${g\to\infty}$ numerically by  simulating the stochastic Schrodinger equation for $N_F=1$. We employ  the technique of Ref.~\cite{cao2019entanglement} for efficient numerical computations with monitored free fermions using the two-point function. 
We give an explicit proof of the validity of the method in~App.~\ref{app:sse-simulation}. 

Starting from a product state of alternating filled and empty sites (i.e. half-filling on average), we study the bipartite entanglement scaling for various values of $\Gamma/\mJ$.
We analyze the scaling of the bipartite entanglement entropy in the states produced by the dynamics at late times, in order to compare with Eq.~\ref{eq:bipartiteentropy} and the surrounding discussion.

We first compute the (trajectory-averaged) von Neumann entropy $S_{\rm vN}(t,L)$ as a function of time after the ``quench''.
For a given system size $L$,
$S_{\rm vN}(t,L)$ eventually plateaus. We define 
$S_{\rm vN}(L)$ by averaging 
$S_{\rm vN}(t,L)$ over $t$ in an interval $[t_{\rm plateau}, t_{\rm max}]$, where  $t_{\rm plateau}$ is determined by eye
and $t_{\rm max}$ is the maximum time we reach in our simulations.
We checked that our results are not sensitive to the choice of $t_{\rm plateau}$ or the time step $\delta t$ used to discretize the stochastic Schrodinger equation.
Raw data is shown in App.~\ref{app:more-numerics}.

\begin{figure}
    \centering
    \includegraphics[width=\linewidth]{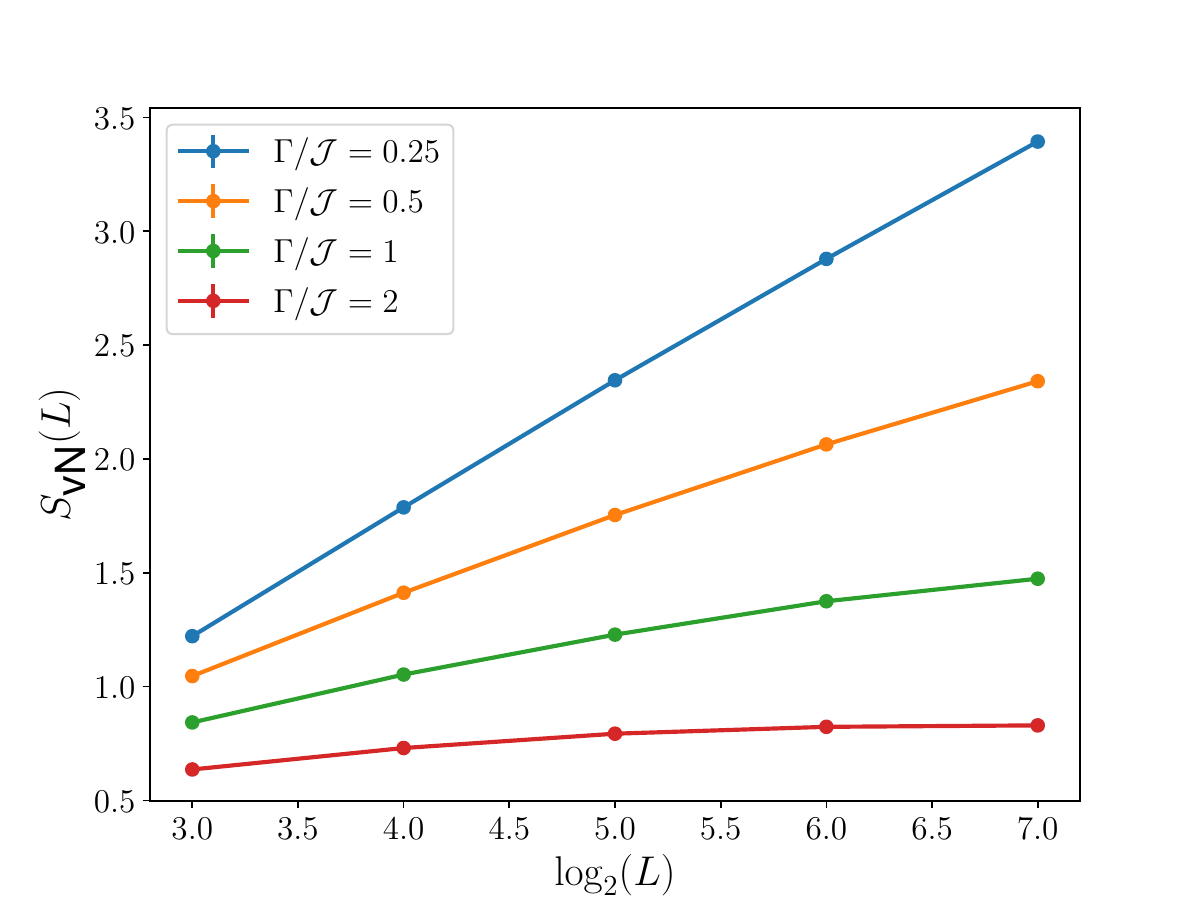}
    \caption{$S_{\rm vN}$ averaged over the time interval $[t_{\rm plateau}, t_{\rm max}]$ (see text). {The statistical errors associated with each dot are not visible on the plot scale.}}
    \label{fig:plateaux}
\end{figure}

In Fig.~\ref{fig:plateaux}, $S_{\rm vN}(L)$ is reported as a function of $\log L$.
Recall that if the steady-state ensemble was scale-invariant, these plots would become straight lines. Our theoretical expectation is that the curves will instead bend downwards, i.e.~will grow sub-linearly, and will give a finite value  for $\lim_{L\to\infty} S_{\rm vN}(L)$. For $\Gamma/\mJ\lesssim 1$ we already see that $S_{\rm vN}$ curves downward, and for $\Gamma/\mJ=2$, $S_{\rm vN}$ reaches a constant value (within error bars) at $L=64$. Physically, this means that the system is in an area-law phase, where entanglement does not grow with the system size.

\begin{figure}
    \centering
    \includegraphics[width=\linewidth]{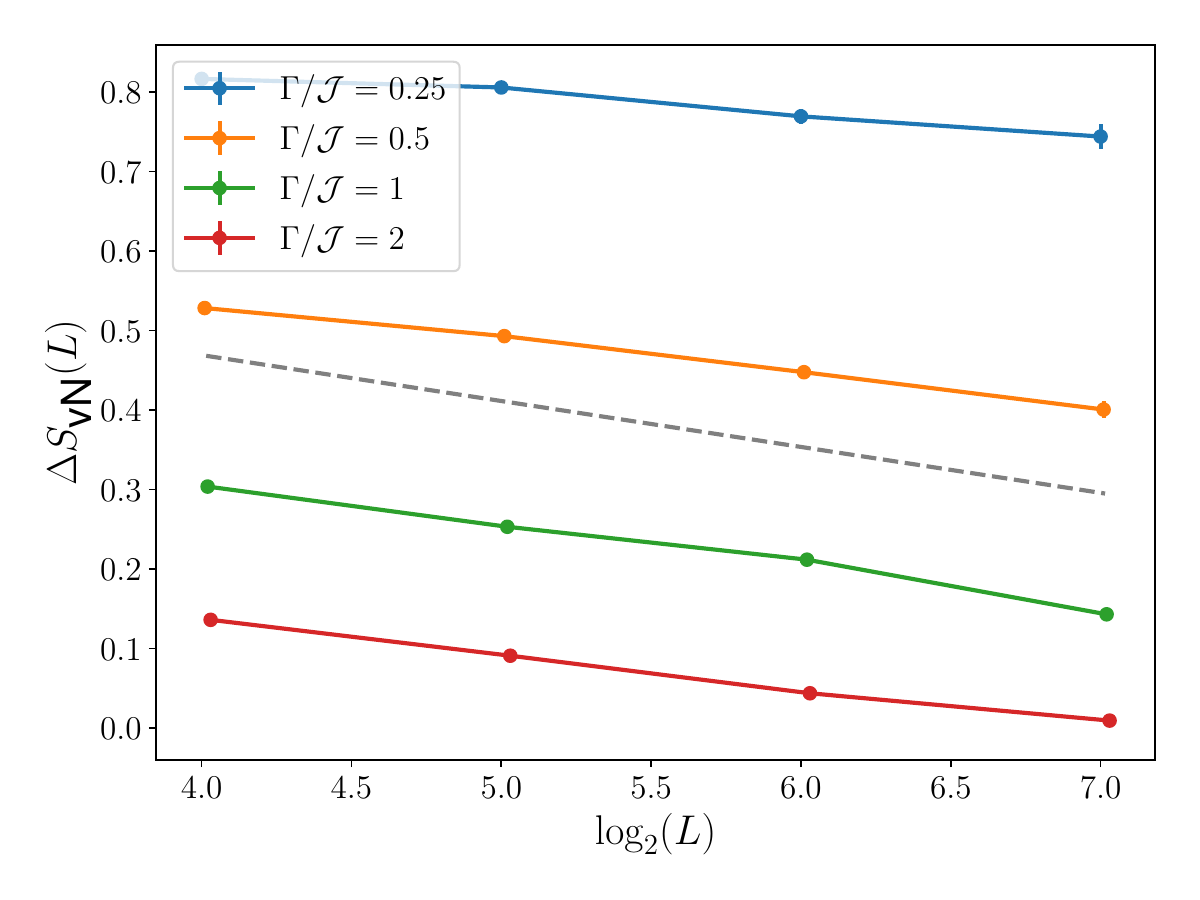}
    \caption{$\Delta S_{\rm vN}$ reported a function of $\log L$. The behavior is approximately linear, in qualitative agreement with~\eqref{eq:bipartiteentropy}. A gray dashed line reports the slope expected from Eq.~\eqref{eq:bipartiteentropy}. The statistical errors associated with each dot are not visible on the plot scale.}
    \label{fig:slope}
\end{figure}

For a more quantitative examination of the curvature, in Fig.~\ref{fig:slope} we report the discrete logarithmic derivative
\be
    \Delta S_{\rm vN}(L) = \f{S_{\rm vN}(L) - S_{\rm vN}(L/2)}{\log 2}
\ee
as a function of $\log L$.
For comparison, Eq.~\ref{eq:bipartiteentropy} predicts that, in the regime (of small renormalized coupling) where $\Delta S_{\rm vN}$ is large,
\be\label{eq:deltaSbehavior}
    \Delta S_{\rm vN}(L) \approx \f{\pi}{3} g_B^{-1} - \f{1}{12} \log L
\ee
plus terms of order 1.
We see that our simulations are not in the regime $g_B\ll 1$, so are far away from the region where the perturbative calculation of $\f{\dd g}{\dd \log L}$ is applicable. Thus, we cannot expect~\eqref{eq:bipartiteentropy} to be accurate. 
Nonetheless, we see that the behaviour is qualitatively similar to the predicted one, and even the observed slope of $\Delta S_{\rm vN}$ is not far from the limiting value $-1/12$ at large $g_B^{-1}$.

To quantitatively check the universal $1/12$ coefficient in Eq.~\ref{eq:deltaSbehavior} it would be necessary to achieve a larger $g_B^{-1}$, either through larger $N_F$, or through smaller $\Gamma/\mJ$. 
The latter requires large length and time scales, because of the crossover from the unitary behavior which obtains for $\Gamma=0$.

\section{Enlarged replica symmetry for fixed-phase hopping}\label{sec:enlargedsymm}

Above we considered models where the phases of the hoppings $J_{j}$ change randomly in time. 
The resulting symmetry class also includes sufficiently generic models with time-independent hoppings.
Formally, this is because both kinds of model have the same replica symmetry.

However, it was appreciated in Ref.~\cite{jian2022criticality} 
in the closely-related setting of non-unitary circuits that imposing constraints on the hoppings can lead to a larger replica symmetry and a different NL$\sigma$M.
(Ref.~\cite{jian2022criticality}
 noted for example that this enlarged replica symmetry applied to the nonunitary circuit in Ref.~\cite{chen2020emergent}.)
To complement the algebraic analysis in Ref.~\cite{jian2022criticality}, we give a simple discussion for a  general free-fermion  Hamiltonian system with monitoring of occupation numbers. Note that enlarged replica symmetry was also discussed in setting of monitored Majorana chains in Ref.~\onlinecite{fava2023nonlinear}.

The  extended replica symmetry holds on a general lattice (in any number of dimensions)  if and only if,
after a gauge transformation of the physical fermion operators,\footnote{A  redefinition  $c_j \rightarrow e^{-i\chi_j} c_j$, with time-independent $\chi_j$.}
all the hopping amplitudes $J_{ij}$ can be made imaginary.
This is equivalent to saying that for every plaquette $P$ of the lattice 
(or generally for every loop made of bonds) the ``gauge flux'' $\phi_P=\sum_{(i,j)\in P} \arg J_{ij}$
takes the value 
$\phi_P = \pi n_{P}$ (modulo $2\pi$), where $n_P$ is the number of edges of $P$.

In particular, this includes all nearest-neighbor chains in which the phase of the hoppings is time-independent, including the models in Refs.~\cite{alberton2021entanglement, turkeshi2022enhanced, coppola2022growth, chen2020emergent, poboiko2023theory}.
Other notable cases are bipartite lattices with real hopping amplitudes~\cite{poboiko2024measurement, chahine2023entanglement}.

When the above condition holds,the effective Hamiltonian commutes with a set of $\mathrm{su}(2N)$ generators even in the presence of measurement, rather than just ${\mathrm{su}(N)\oplus\mathrm{su}(N)}$ as in the more generic setting discussed in the rest of this paper. 
This $\mathrm{su}(2N)$ symmetry becomes apparent after making a particle-hole transformation for the replicas with ${\sigma=-}$, see Appendix \ref{appendix:extended-sym}.

We do not here derive the effective field theory for the case with enlarged replica symmetry, though this could be approached using similar methods to the present ones, for a model with e.g. random imaginary hoppings $J_{ij}$.
However, we can guess the appropriate NL$\sigma$M manifold by considering the action of $\mathrm{SU}(2N)$ symmetry on the replicated infinite-temperature state. This is discussed in App~\ref{appendix:extended-sym} and 
 gives the manifold $\mathrm{SU}(2N)/\mathrm{Sp}(2N)$.
 This NL$\sigma$M manifold is consistent with the symmetry considerations for the ${N\to 0}$ limit in \cite{jian2022criticality}.

Although the NL$\sigma$M on $\mathrm{SU}(2N)/\mathrm{Sp}(2N)$  is different to the one on $\mathrm{SU}(N)$ discussed in the rest of this paper,  
the qualitative features of the renormalization group flow in the limit $N\to 1$ are similar,  
 because although the beta functions of the two models are not identical, both have a positive sign for the $O(g^2)$ term when $N\to 1$.
Therefore we expect a similar physical picture: 
the monitored system is in an area-law phase, 
but if the bare NL$\sigma$M coupling is small there is a semiclassical regime with (loosely speaking) logarithmic scaling of entanglement. 
In this semiclassical regime, the ``entanglement'' degrees of freedom will again be sensitive to the charge density.

\section{Conclusions \& future directions}
\label{sec:conclusions}

In this paper we have given a precise derivation of the effective continuum field theory for monitoring of charge-conserving free fermions.
This derivation illuminates the effect of conserved charge modes (hydrodynamics) on the entanglement entropy.
It also clarifies the symmetry of the nonlinear sigma model, 
which has not been so straightforward to fix in other approaches \cite{poboiko2023theory,poboiko2024measurement,chahine2023entanglement}.

The present approach is controlled by a simple large parameter ($N_F$), and  allows us to take into account the initial states and boundary conditions relevant to computations of entanglement entropies.
Using these results we have given scaling forms that are universal in the ``renormalized semiclassical'' regime  of lengthscales $L\ll \xi$ that are much smaller than the typical lengthscale for entanglement. 
Recall that in 1+1 dimensions $\xi$ can be exponentially large in microscopic parameters such as $N_F$  or the inverse measurement rate  $\Gamma^{-1}$ (and in higher dimensions $\xi$ can be infinite).

The continuum theory derived here  is applicable to essentially any observable that is nonlinear in the density matrix, such as  the statistics of charge fluctuations \cite{poboiko2023theory}, relevant to ``charge sharpening'' \cite{agrawal2022entanglement}.\footnote{We thank S. Gopalakrishnan for discussions of this point.}

More generally, 
the replica field theory approach could be used  to study the detailed fluctuating hydrodynamics of charge in both  unitary and nonunitary noisy dynamics of free fermions. 
In the unitary case this will involve studying Goldstone fluctuations in a Lagrangian analogous to Eq.~\ref{eq:L-ferro} for an $\mathrm{su}(2N)$ ferromagnet \cite{10.21468/SciPostPhys.12.1.042}.

Monitored fermion systems give a new class of phase transitions to explore.
It will be interesting to study the zoology of critical points in ${N\to 1}$ NLSMs both in 1+1 and in  higher dimensions, 
and to explore the analogies/contrasts with critical points both at ${N>1}$ (i.e. conventional ground state phase transitions in spin models \cite{fradkin2013field}) and at ${N\to 0}$ (i.e.  Anderson localization transitions \cite{evers2008anderson} or transitions in random Gaussian tensor networks \cite{jian2022criticality}). 
It would also be worthwhile to develop a simple understanding, for completely general free fermion measurement problems, of how  physical symmetries or constraints on the Hamiltonian and   the monitored operators determine the {\NLSM}  symmetry class \cite{jian2022criticality}. 
See  Sec.~\ref{sec:enlargedsymm} above for the case of U(1)-conserving fermions with measurements of the local density, where we have contrasted generic hopping Hamiltonians from those which give rise to ``real'' Kraus operators and therefore an enlarged replica symmetry, and see Sec.~VIII of \cite{fava2023nonlinear} for the analogous issue in the case of Majoranas.

Another natural question is whether there are generic differences between the $N\to 1$ and $N\to 0$ classes of sigma models that are imposed by the different replica group theory structure. 
For the models discussed here, with target spaces $\mathrm{SU}(N)$ or $\mathrm{SU}(2N)/\mathrm{Sp}(2N)$, it is worth noting that while the ${N\to 1}$ sigma models are ultimately massive in 1+1D, the ${N\to 0}$ sigma models are critical \cite{jian2022criticality, evers2008anderson}.
(This is related to the factor of $N$ in the beta function in  Eq.~\ref{eq:betafunction}.)
While the limit ${N\to 1}$ describes measurements, the limit ${N\to 0}$ arises in problems of forced measurements, whose ``outcomes'' are not sampled with Born's rule, but instead treated as independent random variables. This is why simulations of forced measurements of U(1) fermions show a scale-invariant phase~\cite{chen2020emergent}.

Intuitively, the difference between real and forced measurements, for number-conserved fermions, is consistent with the fact that in the limit of strong continuous-time measurement of  local densities we clearly obtain an area-law state with well-localized fermions \cite{cao2019entanglement}, 
whereas continuous-time \textit{forced} measurement of local densities does not effectively localize fermions. 
This also has a single-particle analog  \cite{jin2023measurement}.
The role played by the charge sector of the effective field theory is also quite different at $N\to 0$ and at $N\to 1$, since it is only in the latter case that this sector has diffusive scaling.

In the future it would be possible to study the effect of out-of-equilibrium charge hydrodynamics on the von Neumann entropy in more detail by  finding solutions to the saddle-point equations\footnote{For more general final-time boundary conditions we must use the  equations in App.~\ref{app:semiclassicaleqmot}. In general we must also use the appropriate scale-dependent \NLSM coupling in solving the saddle-point equations.}  in Eq.~\ref{eq:chargehydro} for, say, nontrivial initial charge profiles or for boundary driving 
\cite{bernard2023exact,PhysRevX.9.021007,rakovszky2019entanglement}.
  The results could then be compared to the effect of charge flow on the mutual information of the Quantum Symmetric Simple Exclusion Process (i.e. in the unitary case) \cite{bernard2023exact}.
As noted above, the regime in which this semiclassical approach is valid can extend over a very large range of scales even in one dimension if $g_B$ is small, and over arbitrarily large scales in higher dimensions.
Since the  dynamical exponent for charge diffusion is 2 --- whereas the natural dynamical exponent for the entanglement sector is 1 ---   
the saddle-point problem may simplify in some settings where we can approximate the charge density  profile as static.

Another interesting topic is the crossover induced by \textit{interactions} between the physical fermions (in fact Gaussianity can be broken either in the Hamiltonian or via the choice of measured operators). 
Adding 4-fermion terms to the physical Hamiltonian breaks the continuous replica symmetry of the entanglement sector down to a discrete symmetry. 
The simplest setting is the Majorana model of Ref.~\cite{fava2023nonlinear}, where the analog of the matrix $Q$ is a real $\mathrm{SO}(N)$ matrix obeying a nonlinear sigma model action  (and there is no charge sector to consider).  
Weak 4-Majorana terms lead to an additional potential which reduces the continuous ground state manifold of the sigma model, $\mathrm{SO}(N)$, to a discrete set. 
It would be interesting to analyze semiclassical  solutions in this model, extending the analysis for the ($N=2$) unitary case in Ref.~\cite{unitarymajorana}. This is a concrete step towards an effective theory for interacting monitored systems.
Note that this route differs from previous approaches \cite{potter2022entanglement,nahum2023renormalization}.

In passing, we note that, while we have focussed on monitored dynamics, 
formally similar problems arise if we measure the physical degrees of freedom in the bulk of a 2D random PEPS wavefunction or similar.
Again the $N\to 1$ limit arises naturally to handle Born's rule and in the free fermion case we expect an NL$\sigma M$ description. 
This context allows for a larger family of tensor networks that do not represent Kraus operators \cite{nahum2023renormalization},
so allows for additional tuning parameters.

Progress could also be made on the formalism for the free fermion dynamics problems. 
In the context of Anderson localization it is well known that an alternative to replicas is the ``supersymmetry'' method \cite{efetov1999supersymmetry}, in which the path integral is formulated using bosons and fermions.
In the present model, the replica spin chain (\ref{eq:Hamiltonian-final}) would become a chain of interacting superspins \cite{zirnbauer1994towards,kondev1997supersymmetry,gruzberg1999exact}.
It would be interesting to see how the separation of the charge and entanglement sectors works in this formalism. This may lead to a more rigorous field-theoretic description of the dynamics. 
The supersymmetric formalism might also give a simple picture for the strongly disentangled limit, in terms of a product ground state of the superspin chain.

Finally, we note that at present there is a lack of mathematically rigorous results for the phase structure of monitored many-body systems (we lack even proofs of the  stability of ``trivial'' disentangled phases).  The formal analogies with Anderson localization  give hope that the free fermion setting may be simpler in this regard than the interacting case. 

\section*{Acknowledgments}
We thank A. De Luca, X. Turkeshi  and S. Gopalakrishnan
for discussions.
DB thanks A. D. Mirlin for interesting discussions. This work was supported by the CNRS, the ENS, the ANR project ESQuisses under contract number ANR-20-CE47-0014-01 and co-funded by the European Union (ERC, QUANTHEM, 101114881). Views and opinions expressed are however those of the author(s) only and do not necessarily reflect those of the European Union or the European Research Council Executive Agency. Neither the European Union nor the granting authority can be held responsible for them.

\appendix

\section{Stochastic formulation of quantum trajectories and their linearisation}
\label{app:stochastic-schroedinger-formulation}

We explain here the link between the $N\to 1$ replica trick and the stochastic formulation of quantum trajectories. The latter are non-linear, quadratic, stochastic differential equations (SDE). They can be linearized at the price of changing the measure and applying Girsanov's theorem in a way similar to the $N\to 1$ replica trick.

Let us start with the simplest case of quantum trajectories with a single monitored operator. For a quantum system with density matrix $\rho_t$ and hamiltonian $H$, and a monitored operator $Q+Q^\dag$, these equations are:\footnote{There is another version with Poisson noise.}
\begin{eqnarray} \label{eq:qtrajec}
d\rho_t &=& -i[H,\rho_t]dt + \mathcal{L}_Q(\rho_t) dt \\
 && + (Q\rho_t+\rho_t Q^\dag - \rho_t \mathrm{tr}(Q\rho_t+\rho_t Q^\dag))dB_t~, \nonumber\\
dY_t &=& \mathrm{tr}(Q\rho_t+\rho_t Q^\dag)dt + dB_t~,
\label{eq:qtrajec-signal}
\end{eqnarray}
with $\mathcal{L}_Q(\rho)=Q\rho Q^\dag - \frac{1}{2}(\rho Q^\dag Q + Q^\dag Q\rho)$ the Lindbladian associated to the so-called jump operator $Q$. Here $B_t$ is a normalized Brownian motion w.r.t. some measure that we denote $\mathbb{E}$, $dB_t^2=dt$  (we shall call a $\mathbb{E}$-Brownian motion a Brownian motion w.r.t. the measure $\mathbb{E}$), and $Y_t$ is the output signal of the ``monitored" (not measured) operator $Q+Q^\dag$.

In the case considered in the main text, $Q$ is Hermitian and proportional to a fermion occupation number, but here we allow for a more general monitoring protocol. For instance, if we would like to monitor the current ${j_k:=i(c_k^\dag c_{k+1} - c_{k+1}^\dag c_k)}$ we could choose $Q=ic_k^\dag c_{k+1}$.

The linearization of those quantum trajectories, and hence the replica trick, amounts to showing that these equations are equivalent to the following linear equations for an un-normalized matrix which we denote by the symbol $\check{\rho}_t$,
\begin{align} \label{eq:sigma-XY}
\rho_t &= \check{\rho}_t/ \mathrm{tr}\check{\rho}_t,
\\
Y_t & = X_t,\\
d\check{\rho}_t &= -i[H,\check{\rho}_t]dt + \mathcal{L}_Q(\check{\rho}_t) dt + (Q\check{\rho}_t+\check{\rho}_t Q^\dag)dX_t,
\label{eq:sigma-linear}
\end{align}
with $X_t$ a normalized Brownian motion w.r.t. to another measure $\mathbb{E}_0$, provided we define the previous measure by 
\begin{eqnarray} \label{eq:newE-E0}
\mathbb{E}[(...)]=\mathbb{E}_0[(\mathrm{tr}\check{\rho}_t)\, (...)]~.
\end{eqnarray}
In particular,
\begin{equation} \label{eq:E-rho-sig}
\mathbb{E}\big[\rho_t\otimes\cdots\otimes\rho_t\big]= \mathbb{E}_0\big[(\mathrm{tr}\check{\rho}_t)\, \frac{\check{\rho}_t}{\mathrm{tr}\check{\rho}_t}\otimes\cdots\otimes   \frac{\check{\rho}_t}{\mathrm{tr}\check{\rho}_t} \big]~,
\end{equation}
as in the setup for the replica trick, where we would go on to write this formula (for the case where there are $k$ factors of $\rho_t$ on the left-hand side) as 
\begin{equation}
\mathbb{E}\big[\rho_t\otimes\cdots\otimes\rho_t\big]= 
\lim_{N\to 1}
\mathbb{E}_0\big[(\mathrm{tr}\check{\rho}_t)^{N-k} \,{\check{\rho}_t^{\otimes k}}\big].
\end{equation}

The proof of Eq.~\ref{eq:sigma-XY}-\ref{eq:newE-E0} is based on Girsanov's theorem (see e.g.~\cite{ksendal2003, Revuz1999}). Let us start with \eqref{eq:sigma-linear} for $\check{\rho}_t$ and recall that $X_t$ is an $\mathbb{E}_0$-Brownian motion. We let $Z_t:=\mathrm{tr}(\check{\rho}_t)$, with $Z_{t=0}=1$, and $M_tZ_t:=\mathrm{tr}(Q\check{\rho}_t+\check{\rho}_t Q^\dag)$. Note that $M_t:=\mathrm{tr}(Q\rho_t+\rho_t Q^\dag)$ with $\rho_t =\check{\rho}_t/ \mathrm{tr}\check{\rho}_t$. We let $\mathcal{L}(\rho) := -i[H,\rho]+ \mathcal{L}_Q(\rho)$.

First (about the change of measure), $Z_t$ is bounded and satisfies $dZ_t= Z_tM_t dX_t$ so that $Z_t$ is $\mathbb{E}_0$-martingale. In particular, $\mathbb{E}_0[Z_t]=1$ is time independent. We deform the measure $\mathbb{E}_0$ by $\mathbb{E}[(\cdots)] = \mathbb{E}_0[(Z_t(...)]$, as in \eqref{eq:newE-E0}. By Girsanov's theorem, $B_t$ defined by $dB_t=dX_t-M_tdt$ is then a $\mathbb{E}$-Brownian motion. That is:
\begin{equation}
 dX_t = M_t dt + dB_t~,
\end{equation}
with $B_t$ a $\mathbb{E}$-Brownian motion (but not a $\mathbb{E}_0$-Brownian motion).

Second (about the SDE),  since $dZ_t= Z_tM_t dX_t$ we have $dZ_t^{-1}= -Z_t^{-1}M_t dX_t+ Z_t^{-1}M_t^2 dt$, via Ito calculus. Define $\rho_t:=Z_t^{-1}\check{\rho}_t=\check{\rho}_t/ \mathrm{tr}\check{\rho}_t$. We have 
\[
d\rho_t=Z_t^{-1}d\check{\rho}_t + (dZ_t^{-1})\check{\rho}_t +  (dZ_t^{-1})d\check{\rho}_t~,
\]
via Ito calculus. Expanding yields
\begin{align}
d\rho_t &= \mathcal{L}(\rho_t)dt 
 + (Q\rho_t+\rho_t Q^\dag)(dX_t- M_tdt) \\
 & ~~~~~~~ -\rho_t M_t (dX_t - M_t dt) ~. \nonumber
\end{align}
Using  $dX_t = M_t dt + dB_t$ and $M_t = \mathrm{tr}(Q\rho_t+\rho_t Q^\dag)$, this reproduces  \eqref{eq:qtrajec} and \eqref{eq:qtrajec-signal}.

The quantum trajectory equations \eqref{eq:qtrajec} and \eqref{eq:qtrajec-signal}, and the statement about their linearization, can clearly be generalized replacing $Q$ by a collection of operators, $Q\to Q_j$, and Brownian motions $dB_t\to dB_t^j$, as in the problem studied in the main text.  It can also by generalized replacing $ -i[H,\rho_t] \to  -i[H,\rho_t] + \mathcal{L}_\mathrm{sys}(\rho_t)$ for some system Lindbladian $\mathcal{L}_\mathrm{sys}$.

\section{Fermionization of the replicas}
\label{app:sec:fermionized-replicas}

In this section, we introduce a replicated Hilbert space for the problem. This issue was already solved in a number of contexts~\cite{10.21468/SciPostPhys.12.1.042, sunderhauf2019quantum, agarwal2022emergent, fava2023nonlinear}, nonetheless, we summarize it here to make the manuscript self-contained and to fix the notation.

Loosely speaking, the replicated is a tensor product of $N$ copies of the Hilbert space of a single chain. However, subtleties arise due the fact that tensor product of fermionic Fock spaces are non-trivial to define. We circumvent this problem by transforming the original fermionic Fock space into a standard Hilbert space, for which replicas can be trivially defined, and, finally, by reintroducing fermionic operators in the replicated Hilbert space.

On a single copy we write the fermionic operators in terms of spin-$1/2$ operators through a Jordan-Wigner transformation.
\begin{align}
\label{app:eq:jordan-wigner}
    c_k &= \Big(\prod_{k'<k} Z_{k'} \Big) S^+_k,
    &
    c_k^\dag &= \Big(\prod_{k'<k} Z_{k'} \Big) S^-_k.
\end{align}
In terms of the bosonic (spin) Hilbert space $\mathcal{H}_B$, we can consider tensor products and map operators to states in a doubles Hilbert space
\be
\label{eq:operator-to-state}
    O = \sum_{i,j} O_{ij} \ket{i}\bra{j} \qquad\mapsto\qquad \ket{O} = \sum_{i,j} O_{ij} \ket{i} \otimes \ket{j}.
\ee
Note that the isomorphism above is basis-dependent, but the actual choice of basis is inconsequential for our treatment.
$N$ replicas of a density matrix can be then identified with states in $\mathcal{H}_{2N} = (\mathcal{H}_B)^{\otimes 2N}$. In this space we can define replicated Pauli operators
\ba
    V^{(+a)}_k &= \mathbb{I}^{\otimes 2 a} \otimes V_k \otimes  \mathbb{I}^{\otimes 2 N - 2 a - 1}\\
    V^{(-a)}_k &= \mathbb{I}^{\otimes 2 a+1} \otimes V_k \otimes  \mathbb{I}^{\otimes 2 (N-a-1)}
\end{align}
with $V= X,\,Y,\,Z$ and $\mathbb{I}$ denoting the identity on the space $\mathcal{H}_B$. Our final aim is to define fermionic operators in this space.
As an intermediate step, we introduce the operators $\chi_{k}^{(\sigma a)}$:
\ba
    \chi^{(+a)}_k &= \mathbb{I}^{\otimes 2 a} \otimes c_k \otimes  \mathbb{I}^{\otimes 2 N - 2 a - 1}\\
    \chi^{(-a)}_k &= \mathbb{I}^{\otimes 2 a+1} \otimes c_k^T \otimes  \mathbb{I}^{\otimes 2 (N-a-1)}
\end{align}
with $c_k^T$ denoting the transpose of $c_k$ w.r.t. the basis used in defining the isomorphism~\eqref{eq:operator-to-state}.
Note that by virtue of this convention we effectively performed a particle-hole transformation on the backward replicas. This will help to make certain symmetries of our model explicit.

Finally, we note that the operators $\chi$ do not anticommute when in different replicas. To fix this, we introduce the Klein factors
\be
    F^{(\sigma a)} = \prod_{k=1}^{L N_F} \exp\left(i\pi \left(\chi^{(+a)}_k\right)^\dag \chi^{(+a)}_k\right)
\ee
and in terms of these the fermionic operators on the replicated Hilbert space
\begin{align}
    c^{(+a)}_k &= i \left( \prod_{a'<a} F^{(+a')} F^{(-a')} \right) F^{(+a)} \chi^{(+a)}_k\\
    c^{(-a)}_k &= \left( \prod_{a'<a} F^{(+a')} F^{(-a')} \right) F^{(+a)} \chi^{(-a)}_k.
\end{align}
It is straightforward to verify that these operators satisfy the canonical anticommutation relation
\be
    \left\{ (c_l^{\alpha})^\dag, c_m^{\beta} \right\} = \delta_{\alpha,\beta} \delta_{l,m}.
\ee

Finally, note that this convention can also be obtained by re-expressing the complex fermions in terms of Majoranas $\gamma_k$, introducing Majorana operators $\gamma^{(\sigma a)}_k$ in the replicated space as in Refs.~\onlinecite{sunderhauf2019quantum,fava2023nonlinear}, and finally writing complex fermions in the replicated space as
\be
    c_j^{(\sigma a)} = \f{\gamma_{2j+1}^{(\sigma a)} - i \gamma_{2j+2}^{(\sigma a)}}{2}.
\ee

\section{Derivation of the spin Hamiltonian}
\label{app:derivation-Hamiltonian}
In this Appendix we derive the effective spin Hamiltonian generating the time evolution $\mathbb{E}_G [(K\otimes K^*)^{\otimes N}]$ in the replicated space.
The starting point is the non-Hermitian Hamiltonian~\eqref{eq:non-hermitian-H-1-replica} generating $K$. In the space with $N$ forward-evolving replicas and $N$ backward-evolving replicas, using the convention detailed in the previous section the operator $(K\otimes K^*)^{\otimes N}$ is generated by
\begin{subequations}
\label{eq:replicated-Hamiltonian}
\ba
H^{(N)}_{\rm meas} &= H_{\rm hop}^{(N)} + H_{\rm flav}^{(N)} + H_{\rm nH}^{(N)} - i \Gamma L N_F N\\
H_{\rm hop}^{(N)} &= \sum_{j,\mu,\nu,\alpha}  J_j^{\mu\nu}(t) c_{j,\mu}^{(\alpha)\dag} c_{j+1,\nu}^{(\alpha)} + {\rm h.c.}\\
H_{\rm flav}^{(N)} &= \sum_{j,\mu,\nu,\alpha} \left( h_{j}^{\mu\nu}(t) c_{j,\mu}^{(\alpha)\dag} c_{j,\nu}^{(\alpha)} - \frac{1}{2}\delta_{\mu\nu} \right)\\
H_{\rm nH}^{(N)} &= i \sum_{j,\mu,\alpha} M_{j,\mu}(t) \sigma_\alpha \left( c_{j,\mu}^{(\alpha)\dag} c_{j,\mu}^{(\alpha)} -  \frac{1}{2} \right).
\end{align}
\end{subequations}

Finally, the Gaussian average over $J$, $h$, and $M$ can be easily obtained through a cumulant expansion, and using the convention detailed in the main text's footnotes, yielding
\ba
N_F \mathcal{H} & = 
\f{\mJ}{2N_F} \sum_{\substack{j,\mu,\nu \\ \alpha,\beta}}(c^{\alpha\dag}_{j\mu}c^\alpha_{j+1,\nu}
c^{\beta\dag}_{j+1,\nu}c^\beta_{j\mu}
+ [j\leftrightarrow j+1])\nonumber
\\
&
+\f{\mathfrak{h}}{2 N_F} \sum_{j,\mu,\nu} \left[
\left(\sum_{\alpha} c^{\alpha\dag}_{j,\mu} c^\alpha_{j,\nu}\right) - N \delta_{\mu\nu}
\right]^2 \nonumber
\\
& 
- \f{\Gamma \sigma_\alpha\sigma_\beta}{2}
\sum_{j,\mu,\alpha,\beta}
c^{\alpha \dag}_{j\mu} c^{\alpha}_{j\mu} 
c^{\beta\dag}_{j\mu} c^\beta_{j\mu} \nonumber
\\
& + \f{\Gamma L N N_F}{2}.
\end{align}
This can be most easily expressed in terms of $S_j$ and $T_j$: the local generators of $\mathrm{SU}(2N)$ and $\mathrm{SU}(N_F)$ rotations respectively:
\ba
S_j^{\alpha\beta}
& =\f{1}{N_F} \sum_\mu c^{\alpha \dag}_{j\mu} c^\beta_{j\mu}
-
\f{\delta^{\alpha\beta}}{2N N_F} \aleph,
\\
T_{j;\mu \nu} & = 
\sum_{\alpha} c^{\alpha\dag}_{j\mu}
c^{\alpha}_{j\nu}
-\f{\delta_{\mu\nu}}{N_F}\aleph,
\end{align}
where we recall that $\aleph$ is the $\mathrm{U}(1)$ generator
\be
    \aleph_j=\sum_{\alpha,\mu} c^{\alpha\dag}_{j\mu} c^\alpha_{j\mu}.
\ee
Then
\ba
N_F \mathcal{H} & = \sum_j \bigg[
- N_F \mJ 
\tr(S_j S_{j+1}) \nonumber\\
&
+\f{\mathfrak{h}}{2 N_F}\lf 
\tr T_j^2 + \f{1}{N_F}(\aleph_j-NN_F)^2 
\ri \nonumber
\\
& 
- \f{\Gamma \sigma_\alpha\sigma_\beta}{2}
\sum_\mu
c^{\alpha \dag}_{j\mu} c^{\alpha}_{j\mu} 
c^{\beta\dag}_{j\mu} c^\beta_{j\mu} \nonumber
\\
&- N_F \mJ \f{\aleph_j \aleph_{j+1}-NN_F(\aleph_j+ \aleph_{j+1}))}{2N N_F^2} \bigg]\nonumber
\\
& + \f{\Gamma L N N_F}{2}.
\end{align}
In the limit $\mathfrak{h}\to\infty$ the conditions $\Tr T_j^2=0$ and $\aleph_j = N_F N$ are enforced. In particular, due to $\Tr T_j^2=0$ each site is projected onto a $\mathrm{SU}(N_F)$ singlet. Therefore in this limit we need the Hamiltonian projected onto this sector. This allows us to simplify the measurement term. Projection is equivalent to Haar-averaging the operator over $\mathrm{U}(N_F)$ transformations $c_{\mu}^{\alpha}\rightarrow \sum_\nu U_{\mu\nu} c_{\nu}^\alpha$ (a site index $j$ is left implicit)
\ba
& \sum_{M, \mu, \nu, \lambda, \kappa} \mathbb{E} \left[ 
U^*_{M \mu}U_{M \nu}U^*_{M \lambda}U_{M \kappa}
\right] 
c^{\alpha \dag}_{\mu} c^{\alpha}_{\nu} c^{\beta\dag}_{\lambda} c^{\beta}_{\kappa} 
\nonumber
\\
&=
\f{1}{N_F+1}\sum_{\mu, \nu}
\lf
c^{\alpha \dag}_{\mu} c^{ \alpha}_{\mu} c^{\beta\dag }_{\nu} c^{\beta}_{\nu}
+
c^{\alpha\dag}_{\mu} c^{\alpha}_{\nu} c^{\beta \dag}_{\nu} c^{\beta}_{\mu}
\ri 
\end{align}
In this way, we finally obtain that, in the sector with $\tr T_j^2 = 0$
and $\aleph_j=NN_F$, the projected Hamiltonian is
\ba
\mathcal{H} = & 
-  \mJ \sum_j \left[ 
\tr(S_j S_{j+1}) -\f{N}{2} \right]\nonumber
\\
& 
- \f{\hat{\Gamma}}{2}
\sum_{j} \left[ \sum_{\alpha,\beta}
\sigma_\alpha \sigma_\beta(S_j^{\alpha\alpha}S_j^{\beta\beta} - S_j^{\alpha\beta}S_j^{\beta\alpha})
+ \f{N}{2}
\right]\nonumber
\\
& + \f{L \Gamma N }{2},
\end{align}
where we introduced ${\hat \Gamma}=\Gamma N_F/(N_F+1)$.

\section{On-site representation}

The aim of this Appendix is to explain why the on-site $\mathrm{su}(2N)$ representation is that with a Young tableau of $N$ rows and $N_F$ columns: for example, in the case $N=3$, $N_F=4$:
\be
{\tiny \ydiagram{4,4,4}}
\ee

First, it is easy to check that $\mathrm{su}(N_F)$ and $\mathrm{su}(2N)$, with generators $S^{\alpha\beta}$ and $T_{\mu\nu}$, are two commuting sub-algebras of $\mathrm{su}(2NN_F)$. 
Recall from the main text that these generators are defined by
\ba\label{eq:Sgenapp}
N_F S^{\alpha\beta} & = \sum_{\mu=1}^{N_F} c_\mu^{\alpha\dag}c_\mu^\beta - \f{\delta^{\alpha\beta}}{2N}\aleph,
\\\label{eq:Tgenapp}
T_{\mu\nu} & = \sum_{\alpha=1}^{2N} c^{\alpha\dag}_\mu c^\alpha_\nu - \f{\delta_{\mu\nu}}{N_F} \aleph,
\end{align}
with total fermion number $\aleph = \sum_{\mu}\sum_\alpha c^{\alpha\dag}_\mu c^\alpha_\mu = N N_F$ in the sector of interest.

We are interested in the $\mathrm{su}(2N)$ representation which is the $\mathrm{su}(2N)$ orbit of the replicated identity map 
(replicated $N_F$ times), 
which in our notation is denoted as a \textit{state}, $\ket{\mathbb{I}}$. 
It is easy to check that it is made of $\mathrm{su}(N_F)$ scalars. 

For $N_F=1$, 
the identity belongs to the $\mathrm{su}(2N)$ irreducible representation of anti-symmetric $N$-tensors, with Young tableau made of one column of length $N$.\footnote{\label{footnote:antisymmetrictensor} Denoting the fermion vacuum (which is manifestly invariant under Eqs.~\ref{eq:Sgenapp},~\ref{eq:Tgenapp}) by $\ket{\Omega}$,
the sector with $\aleph=NN_F$ is given by filling $N N_F$ fermion modes. 
For $N_F=1$, this gives states $A_{\alpha_1, \ldots, \alpha_N} c^{\alpha_1 \dag}\cdots c^{\alpha_N\dag} \ket{\Omega}$, where $A$ is an antisymmetric tensor.} Within this $\mathrm{su}(2N)$ representation there is a state $v$ which is a $\mathrm{su}(2N)$ highest weight vector (for the given selected Borel sub-algebra). 

For $N_F>1$, the $\mathrm{su}(2N)$ action is the diagonal one on the $N_F$-fold tensor product of the above $N_F=1$ representation. The state $v\otimes\cdots\otimes v$ ($N_F$ times) is thus in the orbit of the replicated identity under the $\mathrm{su}(2N)$ action. Because the $\mathrm{su}(2N)$ action is the diagonal one, it is also an $\mathrm{su}(2N)$ highest weight vector. Its highest vector is $N_F$ times that of $v$. Hence it is the highest weight vector of the representation with Young tableau made of $N_F$ column and $N$ lines. That is: the irreducible $\mathrm{su}(2N)$ representation containing the identity map is that with Young tableau made of $N_F$ column and $N$ lines. 

If one is not at ease with this general argument, one may check it on the first few cases. For $N_F=2$,  $2N=2$:  the sector with ${\aleph=2}$ has dimension $6$, and is made of $\mathrm{su}(4)$ anti-symmetric 2-tensors.\footnote{  Similarly to the discussion in Footnote~\ref{footnote:antisymmetrictensor},
this sector is obtained by filling ${\aleph=2}$ fermion modes, so contains states of the form
$A_{IJ}c^\dag_I c_J \ket{\Omega}$, where
$I,J$ are multi-indices running over $2NN_F$ values, and $A$ is an antisymmetric tensor.}
Projecting it on the $\mathrm{su}(N_F)$ invariant subspace yields a vector space of dimension $3$ which is an irreducible triplet w.r.t. $\mathrm{su}(2N)$ (recall $2N=2$). That is: in this case, fixing $\aleph=NN_F$ and projecting on $\mathrm{SU}(N_F)$-scalars selects an $\mathrm{su}(2N)$ irreducible representation. One can check directly (on a small piece of paper) that this also works for $N_F=3, 2N=2$ as well as for  $N_F=2, 2N=4$. In all these cases the resulting $\mathrm{su}(2N)$ irreducible representation is that with rectangular Young tableau with $N$ lines and $N_F$ columns. It is also easy to prove along these lines that the identity belongs to this representation and that it is part of the $\mathrm{su}(N_F)$ invariant with $\aleph=NN_F$. We can thus restrict to it.

\section{From the Heisenberg equations to the NL$\sigma$M action}
\label{app:Heisenberg-equations}

As explained in the main text, the Hamiltonian $H_N$ has modes associated to $L$ and $R$ that have an explicit mass, and modes associated to the slow variation of $Q$ and (in the $N\to1$ limit) of $d$ and $q$ that are  Goldstone modes in the ultraviolet.

Our aim is then to integrate out $L$ and $R$ to derive an effective Lagrangian $\mathcal{L}[Q, d, q]$ for the other modes. To do this we will exploit the semiclassical limit $N_F\gg 1$.
Here we can integrate out the gapped modes at the level of the equations of motion and then reconstruct the Lagrangian from those.

We define the operators in the Heisenberg picture (recall from the main text that we Wick rotate $t=i\tilde t$ so that the equations of motion have the conventional form)
\ba
\label{eq:def-real-time}
    \dot{\mathcal{O}} &= i N_F [\Hs, \mathcal{O}],
\end{align}
so that on a finite system it gives
\ba
    \mathcal{O}(\tilde t) &= e^{i N_F \Hs \tilde t} \mathcal{O} e^{- i N_F \Hs \tilde t}.
\end{align}
We find that, at the operator level, the blocks of $S_j$,
\ba
    S_j &=
    \begin{pmatrix}
        {\tilde L}_j  & {\tilde Q}_j\\
        {\tilde Q}_j^\dag & {\tilde R}_j  
    \end{pmatrix},
\end{align}
satisfy
\begin{subequations}
\ba
    \label{eq:tilde-Q-dot}
    -i\dot{\tilde{Q}}_j =&  \f{\mJ}{2} \left[ \tilde{L}_j  (\tilde{Q}_{j-1}+\tilde{Q}_{j+1}) - (\tilde{Q}_{j-1}+\tilde{Q}_{j+1})  \tilde{R}_j \right]\nonumber\\
    -&\f{\mJ}{2} \left[  (\tilde{L}_{j-1}+\tilde{L}_{j+1}) \tilde{Q}_j - \tilde{Q}_j (\tilde{R}_{j-1}+\tilde{R}_{j+1}) \right] \nonumber\\
    +& 2\hat{\Gamma} (\tilde{L}_j \tilde{Q}_j \!- \!\tilde{Q}_j \tilde{R}_j) \!- \!\hat{\Gamma} \left\{\tr\left( \tilde{L}_j\!-\!\tilde{R}_j\right), \!\tilde{Q}_j \!\right\},\\
    \label{eq:tilde-L-dot}
    -i \dot{\tilde{L}}_j &=    \f{\mJ}{2} \left[ \tilde{Q}_j(\tilde{Q}_{j-1}^\dag + \tilde{Q}_{j+1}^\dag) 
    - (\tilde{Q}_{j-1} + \tilde{Q}_{j+1})  \tilde{Q}_j^\dag \right]\nonumber\\
    -&\f{\mJ}{2} \left[ (\tilde{L}_{j-1} + \tilde{L}_{j+1}) \tilde{L}_j - \tilde{L}_j (\tilde{L}_{j-1} + \tilde{L}_{j+1}) \right],\\
    \label{eq:tilde-R-dot}
    -i \dot{\tilde{R}}_j =&  \f{\mJ}{2} \left[ \tilde{Q}_j^\dag (\tilde{Q}_{j-1} + \tilde{Q}_{j+1}) - (\tilde{Q}_{j-1}^\dag + \tilde{Q}_{j+1}^\dag)  \tilde{Q}_j \right]\nonumber\\
    -&\f{\mJ}{2} \left[ (\tilde{R}_{j-1} + \tilde{R}_{j+1}) \tilde{R}_j - \tilde{R}_j (\tilde{R}_{j-1} + \tilde{R}_{j+1}) \right].
\end{align}
\end{subequations}
Let us parametrize $\tilde Q$ as $\tilde Q = q Q$ with ${\det Q=1}$ and $\tilde L = L+d\mathds{1}$ and $\tilde R = R-d \mathds{1}$ with $\tr L = \tr R =0$. In the continuous limit, the first equation then becomes
\begin{align}
   -i(\dot q + \dot Q Q^{-1}) =&  \mJ \left[ d \partial_x^2(qQ)\, Q^{-1} - q\partial_x^2 q\right]    \nonumber \\
   & + 4\hat \Gamma\, L +4\hat \Gamma (1-N)\, d ~.
\end{align}
where we neglect  terms of order $N_F^{-1}$ 
(arising from order--$N_F^{-1}$ corrections to the kinematic constraints in Eq.~\ref{eq:LQRconstraints})
or $O(\tilde L\partial_x^3 \tilde Q,\tilde R\partial_x^3 \tilde Q, \tilde Q\partial_x^3\tilde L,\tilde Q\partial_x^3\tilde R)$.
Throughout, we organize terms by the number of spatial derivatives, i.e. by the order in $\partial_x$.

We now extract the trace and the traceless part of the above equation. The condition ${\det Q=1}$ implies that ${\tr(Q^{-1}\delta Q)=0}$ for any infinitesimal variation. In particular $Q^{-1}\dot Q$ and $Q^{-1}\partial_x Q$ are traceless. Thus we get
\begin{subequations}
\begin{align}
    -iN\dot q  = & \, \mJ\big[ N(d \partial_x^2q - q\partial_x^2 q)+dq\tr(\partial_x^2Q\, Q^{-1})\big]  \nonumber\\
    &\, +4\hat \Gamma (1-N)N\, d ~, \\
 -i\dot Q Q^{-1}= & \, + 4\hat \Gamma\, L + \mJ\, d \mathcal{A}~, \label{eq:tracefulpart}
\end{align}
\end{subequations}
with
\begin{align}
 \mathcal{A} = \partial_x^2(qQ)\,Q^{-1} - \frac{\mathds{1}}{N}\tr(\partial_x^2(qQ)\,Q^{-1})~.
 \nonumber
\end{align}
In Eq.~\ref{eq:tracefulpart} 
we neglect $\mathcal{A}$ compared to $L$, 
because it is a higher-derivative contribution.
To see this: the equation below for $\dot L$ shows that $\dot L$ scales as $O(\partial_x^2)$. Therefore $L$, without the time derivative, is much larger than $O(\partial_x^2)$, which is the order of $\mathcal{A}$.

We thus get
\begin{subequations}
\begin{align}\notag
    -iN\dot q  = & \mJ
   \left[ 
    N(d \partial_x^2q - q\partial_x^2 d)+dq\tr((\partial_x^2Q)\, Q^{-1}) \right]  \\
    &+4\hat \Gamma (1-N)N\, d ~,\\
 -i\dot Q Q^{-1}= &  4\hat \Gamma\, L ~,\label{eq:Qdoteqnfinalapp} 
\end{align}
\end{subequations}

Note that since $L$ is hermitian and traceless this last equation is compatible with $Q$ being unitary, $QQ^\dag=\mathds{1}$.
In other words, the long-wavelength fluctuations of $Q$ preserve the unitarity of $Q$ that is a property of the classical ground states. 
Therefore we now impose unitarity.

We get the evolution equation for $L$ and $d$ by looking at Eq.~\ref{eq:tilde-L-dot}. This yields
\begin{equation}
-i(\dot L +\dot d) = \frac{\mJ}{2}[ qQ\partial_x^2(q^*Q^\dag)-\partial_x^2(qQ) Q^\dag q^*] ~.
\end{equation}
Extracting the trace and the traceless part gives
\begin{align}
i\dot d =& \frac{\mJ}{2}(q^*\partial_x^2 q- q \partial_x^2 q^*) ~,\\
i\dot L =& -\frac{\mJ}{2}\Big[ |q|^2 \lf Q\partial_x^2Q^\dag-(\partial_x^2 Q) Q^\dag \ri \nonumber \\
& \quad \quad \, \, \, + 2\lf (q\partial_xq^*)\, Q\partial_xQ^\dag - \mathrm{h.c.}\ri \Big] ~.\label{eq:Ldoteqnapp}
\end{align}
We may combine Eqs.~\ref{eq:Qdoteqnfinalapp},~\ref{eq:Ldoteqnapp} to get the equation for $\ddot Q$ in the main text:
\be
\label{eqs:entanglement-eom-app}
Q^\dag \left[
\ddot Q - 4 \mathcal{J}\hat \Gamma \, 
\partial_x  \big( |q|^2 \partial_x Q \big) 
\right]  - \rm{h.c.} = 0 ~.
\ee
We may write the equation for $q$ in the form 
\ba
  {i} \dot{q} & =  \!  \mJ \! \left(  q\partial_x^2 d -  d \partial_x^2 q  \right) 
    \!+\! q d \left[ 
    \f{\mJ }{N}\!  \tr { \partial_x Q^\dag \partial_x Q }           
    - 4\hat{\Gamma}_N 
    \right]\!
    \end{align}
by using the fact ${\tr(Q^{-1}\partial_x Q)=0}$, together with the unitarity of $Q$, which imply
\[
\tr(\partial_x^2Q\, Q^\dag)=
-\tr(\partial_xQ\, \partial_xQ^\dag)
= \tr (Q\, \partial_x^2 Q^\dag)
\]
and similarly for the $\tilde t$-derivatives.

One can easily check that the equations of motion \eqref{eqs:entanglement-eom} and~\eqref{eqs:charge-eom}  are reproduced by the action~\eqref{eq:action-N}. 
First, one needs to impose that the action is extremal w.r.t. variations of the form $Q\mapsto Q(1+ih)$, with $h=h^\dag$ and $\tr h =0$. 
For this purpose, note that, if $\tr(X h)=0$ for all such $h$, then $X\propto \mathds{1}$. 
The local variation of $S_{\rm NL\sigma M}$ is of the form $\tr (Xh)$ with
\be
X \propto 
Q^\dag \left[
\partial_{\tilde t} \left( \f{1}{g_B v} \dot Q \right) 
-
\partial_{x} \left( \f{v}{g_B}  Q' \right) 
\right] - {\rm h.c}. 
\ee
By the identities above, this is traceless, so it must vanish, giving the equation of motion $X=0$.
Finally, substituting in Eq.~\ref{eq:gandv} for $g_B$ and $v$ as functions of $|q|$ recovers the equation of motion (\ref{eqs:entanglement-eom}) in the main text.

Finally, in order to extremize the action in the main text w.r.t. $\vec{n}$, it is useful to note that
\be
    \delta \left[ \int \! \! dx dt (1-n_z) \dot{\varphi} \right] = \int \! \! dx dt\, \delta \vec{n} \cdot \left( \dot{\vec{n}} \times \vec{n} \right).
\ee
    
\section{Semiclassical equations of motion}
\label{app:semiclassicaleqmot}

In this appendix we discuss the saddle-point solutions contributing to the amplitude
\be
    \braket{\rm fin| e^{-N_F t \Hs}| \rm ini}
\ee
where $\ket{\rm ini}$ and $\bra{\rm fin}$ are determined by the initial density matrix of the system and the observable being computed respectively.

The saddle-point solutions are given by the semi-classical equations of the action~\eqref{eq:action-N}.
In general there will be a coupled set of equations for both the charge degrees of freedom $\vec{n}$ and the entanglement degrees of freedom in $Q$. However we argue below that, when computing the Von Neumann entropy using the $N\to 1$ limit, one can neglect the feedback of $Q$ on the mode $\vec{n}$.

The saddle-point solutions for $Q$ are similar to the ones discussed in Ref.~\onlinecite{fava2023nonlinear}. Here the final and initial state specify two time-like boundary conditions for $Q$, and, given that the equations of motion for $Q$ are second-order in time, this uniquely specifies a solution. 
The solution for the charge sector $\vec{n}$ can be more subtle since it involves boundary conditions at both initial and final times, despite the fact that   the equations of motion are only first-order in time \cite{stone2000semiclassical, tailleur2008mapping,
unitarymajorana}. 
The resolution to the apparent paradox is that the saddle-point equations for the charge sector involve two independent fields (as will be clear below): one  has its boundary condition imposed at the final time, and the other has its boundary condition imposed at the initial time.

It is convenient to represent $\vec{n}$  through stereographic projection of $\vec{n}$ from the point $(0,0,1)$ onto the plane $\{(x,y,-1)|x,y\in\mathbb{R}\}$, so that
\be\label{eq:ndefnzzbar}
    \vec{n} = \f{1}{1+z \bar{z}}
    \begin{pmatrix}
        z+\bar{z}\\
        i (\bar z- z)\\
        1- z \bar{z}
    \end{pmatrix}.
\ee
This corresponds to the coherent states representation of the spins as 
\be\label{eq:spinhalfcoherentstate}
    \ket{z} = \f{1}{\sqrt{1+\bar{z} z}}
    \begin{pmatrix}
    1\\
     z   
    \end{pmatrix},
\ee
in the case of spin-1/2 (i.e. for $N=N_F=1$) 
or as  \cite{stone2000semiclassical}
\be
\label{eq:spincoherentstate}
    \ket{z} \propto
    \exp(z S_-) \ket{S,+S}
\ee
for a general spin $S$ (where the ket on the right-hand-side is the state with maximum $n_z$).

Substituting this into $\mathcal{L}$, we find that, up to a constant, the Lagrangian  (the convention for the path integral weight is $e^{-\int \dd t \dd x \mathcal{L}}$) 
has the form $\mathcal{L} = \f{N N_F}{2}\mathcal{L}_1$ with
\ba
    \mathcal{L}_1  & =   \f{\bar{z} \partial_t z - z \partial_t \bar{z}}{1 + z \bar{z}} 
    +   \mJ   \f{(\partial_x \bar{z}) \partial_x z}{(1 + z \bar{z})^2} + G(Q) \left(\f{1 - z \bar{z}}{1 + z \bar{z}}\right)^2
    \nonumber
    \\
    &+ \text{terms independent of $z,\bar z$}, 
\end{align}
with
\be
    G(Q) = \hat{\Gamma}_N - \f{\mJ}{4N} \tr\left((\partial_x Q^\dag) \partial_x Q \right).
\ee
Extremizing the action w.r.t. $\bar z$ and ${z}$ we get respectively
\ba
  \label{eq:z-eom}
    \partial_t z &= \f{\mJ}{2} \left( \partial_x^2 z
    - 2 \f{\bar{z} (\partial_x z)^2}{1+z \bar{z}} \right) 
    + 2 G(Q) \bar{z} z^2 \f{1-z \bar{z}}{1+z\bar{z}},
    \\
    \partial_t \bar{z} &= -\f{\mJ}{2} \left( \partial_x^2 \bar{z}
    - 2 \f{z (\partial_x\bar{z})^2}{1+z \bar{z}} \right) - 2 G(Q) z \bar{z}^2 \f{1-z \bar{z}}{1+z\bar{z}}.
\end{align}

When $N=1$, the term $G(Q)$ vanishes and the equations simplify: 
\begin{subequations}
\ba\label{eq:zeqsimplified}
    \partial_t z &= \f{\mJ}{2} \left( \partial_x^2 z - \f{2 \bar{z} (\partial_x z)^2}{1+\bar{z} z} \right)\\
    \partial_t \bar{z} &= - \f{\mJ}{2} \left( \partial_x^2 \bar{z} - \f{2 z (\partial_x \bar{z})^2}{1+\bar{z} z} \right).
\end{align}
\end{subequations}
When searching for saddle points of the action, one must then treat $z$ and its complex conjugate $\bar{z}$ as independent variables satisfying the equations of motion.

We expect that, in order  to study the $N\to 1$ limit for the von Neumann entropy,
it is sufficient to study the charge sector only at $N=1$ (instead of having to find the solution for any $N$). 
Therefore below we will focus on Eqs.~\ref{eq:zeqsimplified}, together with the the boundary conditions on $z$, $\bar z$ that arise at $N=1$.

We give a non-rigorous self-consistent argument for the fact that it is sufficient to study the charge-sector equations only for $N=1$. 
In order to obtain the von Neumann entropy,  we need the free energy of the field theory up to order ${N-1}$ (see e.g. Eq.~57 in \cite{fava2023nonlinear}, taking  $k=1$, or \cite{jian2020measurement}).
Note that  the saddle-point solution for $Q$ discussed above satisfies $\tr (\partial_x Q^\dag) (\partial_x Q) \propto {N-1}$ at small $N-1$.
Therefore we expect that the term $G(Q)$ may be treated as formally of order $N-1$, and that 
dropping this term also affects the saddle-point solutions only at relative order $N-1$ compared to the leading order. Therefore it affects the saddle-point action by at most order $(N-1)^2$.
It would be worthwhile to make this argument more precise.

Using coherent states one can show that the boundary states impose non-trivial boundary conditions on $z$ and $\bar{z}$. When the boundary states are coherent states themselves $\ket{\rm ini}= \ket{z_{\rm ini}(x)}$ and $\ket{\rm fin} = \ket{z_{\rm fin}(x)}$, then~\cite{stone2000semiclassical}
\ba
\label{eq:z-boundary-conditions}
    z(x,0) &= z_{\rm ini}(x),
    &
    \bar{z}(x,t) &= \bar{z}_{\rm fin}(x).
\end{align}

When $N=1$,  
the boundary states are coherent states of the form (\ref{eq:spincoherentstate}).
The final-time boundary conditions we consider 
reduce when $N=1$ to the state ``$\bra{\mathbb{I}}$'', and give
$\bar{z}(x,t_{\rm fin})=i$ (see App.~\ref{sec:boundar_states}).
It then follows from Eq.~\ref{eq:zeqsimplified} that   $\bar{z}(x,t)=i$ for all $x$ and $t$, i.e. it is a constant.

We consider physical initial density matrices that are  product states, where each site has a fixed chemical potential $2\mu_j$:
\be \label{eq:chempot}
    \rho = \prod_j \f{e^{\mu_j n_j}}{e^{\mu_j}+1}
\ee
(recall that $n_j = N_F^{-1} \sum_\mu c_{j\mu}^\dag c_{j\mu}$).
For these states we find
(see App.~\ref{sec:boundar_states}) the initial condition (switching to continuum notation)
\be
\label{eq:z-initial-condition1}
z(x,0) = -i f(x),
\ee
where
\be
\label{eq:z-initial-condition2}
     f(x) = e^{-\mu(x)}
\ee
Writing \eqref{eq:z-eom} in terms of $f(x,t)=iz(x,t)$
[and using  $\bar{z}(x,t)=i$] we see that $f(x,t)$ remains real at all times.  
While the equation of motion for $f(x,t)$ does not have a simple form, this reduces to a standard diffusion equation for the physical density of fermions by writing 
\be
    d=\f{1}{2} \f{1-f}{1+f}. 
\ee
which gives
\be
    \partial_t d = \frac{\mJ}{2} \partial_x^2 d.
\ee
Recall that ${d = n-1/2}$, where $n$ is the charge density.

\section{Further properties of boundary states}
\label{sec:boundar_states}

The simplest final state we will consider is $\bra{\mathbb{I}}$, which is used to compute traces, i.e. $\braket{\mathbb{I}|\rho^{\otimes N}}= \Tr(\rho)^N$.
This satisfies $\bra{\mathbb{I}} V^{(+a)} = \bra{\mathbb{I}}  (V^{(-a)})^T$ 
for any Pauli operator $V$ 
in the bosonic representation obtained by Jordan-Wigner
(see App.~\ref{app:sec:fermionized-replicas}). From this we have
\ba
    \bra{\mathbb{I}} c_{j,\mu}^{(-a)} &= -i \bra{\mathbb{I}} c_{j,\mu}^{(+a)}
    &
    \bra{\mathbb{I}} (c_{j,\mu}^{(-a)})^\dag &= -i \bra{\mathbb{I}} (c_{j,\mu}^{(+a)})^\dag,
\end{align}
and thus
\ba
    \bra{\mathbb{I}} S_j = \bra{\mathbb{I}} 
    \begin{pmatrix}
        \tilde{L}_j & i (\f{\mathds{1}}{2} - \tilde{L}_j)\\
        -i (\f{\mathds{1}}{2} + \tilde{L}_j) & -\tilde{L}_j
    \end{pmatrix}.
\end{align}
From this, one can verify the condition stated in the main text that $\bra{\mathbb{I}}S_j^2=\bra{\mathbb{I}}\mathds{1}/4 + O(1/N_F)$.
 Taking the expectation value we have (for any value of $N_F$)
\ba\label{eq:finalstateidentityexp}
    \bra{\mathbb{I}} S_j 
    \ket{\mathbb{I}} 
    = \bra{\mathbb{I}} 
    \begin{pmatrix}
        0 & i \f{\mathds{1}}{2} \\
        -i \f{\mathds{1}}{2}  & 0
    \end{pmatrix}
    \ket{\mathbb{I}} .
\end{align}
When $N=1$, this gives
the expectation value of the SU(2) spin operators as
${(n_x, n_y, n_z) = (0, 1, 0)}$
so that at $N=1$ this is an SU(2) coherent state with $z=-i$ (by  Eq.~\ref{eq:ndefnzzbar}).

From Eq.~\ref{eq:finalstateidentityexp} we also we read off the final-time boundary condition
\be
    Q(x,t_{\rm fin})  = \mathds{1}
\ee
corresponding to the boundary state $\bra{\mathbb{I}}$.

To compute the average Von Neumann entropy of a state, we further need to consider the final state $\bra{{\rm vN}}$ defined by $\braket{\rm vN | \rho^{\otimes N}}= \Tr(\rho^N)$.
The state must then satisfy $\bra{\rm vN } V^{(-a)} = \bra{\rm vN }  (V^{(+(a+1))})^T$ for $a<N$ and $\bra{\rm vN } V^{(-N)} = \bra{\rm vN }  (V^{(+1)})^T$.
While the expectation values of $\tilde{L}_j$ and $\tilde{R}_j$ are zero,
\be
    \braket{\rm vN| \tilde{Q}_j |\rm vN} = \f{i}{2} { Q_{\rm vN}},
\ee
with
\be
{ Q_{\rm vN}} \equiv 
\lf \begin{array}{ccccc}
    \phantom{+} 0 & \phantom{+} 0 & \cdots &  \phantom{+} 0 &  + 1  \\
    -1 & \phantom{+} 0 & \cdots &  \phantom{+} 0 &  \phantom{+} 0  \\
    \phantom{+} 0 & -1 & \cdots &  \phantom{+} 0 & \phantom{+}  0  \\
     &   & \cdots & &     \\
     \phantom{+} 0 & \phantom{+} 0 & \cdots & -1 &\phantom{+}  0
\end{array}\ri.
\ee
The eigenvalues of the matrix $q_n$ are studied in App. I of Ref.~\onlinecite{fava2023nonlinear}. From that we find that $q = i/2$ and $Q = q_n$
In the $N=1$ limit, this again reduces to a state polarized along $y$, fixing $\bar{z}(x,t)=i$.
{ Summarizing: to compute the bipartite entanglement entropy, it is sufficient to impose the boundary condition $Q=\mathds{1}$ on one subsystem and $Q={ Q_{\rm vN}}$ on the other.

Finally, the most generic initial states we will consider are product states where each site has a fixed chemical potential $\mu_j$, as in Eq.~\ref{eq:chempot}.
Taking replicas of this physical state gives the initial state 
\be
    \ket{\rm ini} = \left[ \prod_j \f{\exp\left(\mu_j N_F^{-1}\sum_\mu \sum_{a=1}^N c_{j\mu}^{(+a)\dag} c_{j\mu}^{(+a)}\right)}{(e^{2\mu_j}+1)^{N/2}} \right]\ket{\mathbb{I}},
\ee
where the normalization is fixed by $\braket{\rm ini| \rm ini}=1$.
Taking the expectation valu of the spin operator $S$ on $\ket{\rm ini}$ gives
\ba
    \braket{\rm ini|\tilde{Q}_j|\rm ini} &=  \f{i}{2 \cosh \mu_j} \mathds{1},
    &
    \braket{\rm ini| d_j| \rm ini} &= \f{1}{2}\tanh \mu_j,
\end{align}
while the expectation value of $L$ and $R$ are zero.
One can check that this corresponds to a fully polarized state for $\vec{n}$.
This implies that when $N=1$ it is a coherent state~\eqref{eq:spincoherentstate} with $z=-i e^{-\mu_j}$, fixing the initial condition for $z$ as in~\eqref{eq:z-initial-condition1} and~\eqref{eq:z-initial-condition2}.
}

\section{Proof of the validity of the QR decomposition in simulating the stochastic Schrodinger equation}
\label{app:sse-simulation}

{
Simulating the stochastic Schrodinger equation associated to the time evolution we consider requires discretizing time into intervals of size $\delta t$. For each time step, the Kraus operator is of the form
\be
    K(\delta t) = \exp\left[\sum_j n_j \bigg(\delta B_j + (2\langle n_j \rangle-1) \Gamma \delta t)\bigg)\right] e^{-i H(t) \delta t}
\ee
where $\mathbb{E}[\delta B_j^2]=\Gamma \delta t$.

We are interested in pure initial states. We first consider the case where the system is initialized in 
\begin{equation}\label{eq:dw_initial_state}
    \ket{\psi_0}= c^{\dagger}_1\cdots c^{\dagger}_N\ket{0}\,.
\end{equation}
Namely, all fermions occupy the first $N$ sites in the chain. Later, we will show how to generalize the discussion for more general Fock states, including the so-called Néel state, featuring alternating filled and empty sites. We represent the state of the evolved system as
\begin{equation}\label{eq:state}
	\left|\psi(t)\right\rangle=\mathcal{U}(t) c^{\dagger}_1\cdots c^{\dagger}_N\ket{0} =\prod_{k=1}^{N}\left[\sum_{j=1}^{L} u_{j k} c_{j}^{\dagger}\right]\ket{0}\,,
\end{equation}
where $\mathcal{U}=\exp\left[ -i h_{a,b}(t) c^{\dagger}_a c_b \right]$ and $h_{a,b}(t)$ is an Hermitian matrix. [Here and in the following we use the convention that calligraphic operators act in the many-body space, while lower-case letter operators are the corresponding single-particle-sector $L\times L$ matrices.]

Updating the matrix $u(t)$ after the application of the first time-evolution step $e^{-i H(t) \delta t}$ is trivial.  In contrast, efficiently computing how the non-unitary transformation in the second step affects the covariance matrix is non-trivial and was first discussed in Ref.~\cite{cao2019entanglement}.
In this section we summarize the technique and explicitly prove its validity for pure states. 

The covariance matrix of~\eqref{eq:state} is
\begin{equation}
	\braket{\psi|c^{\dagger}_jc_k|\psi}=u
	\begin{pmatrix}
		\openone_N & 0 \\
		0 & 0 
	\end{pmatrix} u^\dagger\,.
\end{equation}
}

Our goal is to find the exact covariance matrix corresponding to the normalized state
\begin{equation}
	\ket{\psi^\prime}=\frac{1}{\sqrt{\bra{\psi}\mathcal{V}^\dagger \mathcal{V}\ket{\psi}}}\mathcal{V}\ket{\psi}\,,
\end{equation}
where $\mathcal{V}=e^{\varepsilon c^{\dagger}_j c_j}$. Here, $\varepsilon\in \mathbb{R}$ does not need to be small. This can be done using the general method of Ref.~\cite{bravyi2004lagrangian}. 
However, since the state is pure and particle number is conserved, we can use an even more efficient method, which is the QR decomposition put forward in Ref.~\cite{cao2019entanglement}. We explain it explicitly in the following.

First, noticing
\begin{equation}
	\mathcal{V}\ket{0}=\ket{0}\,,
\end{equation}
we can rewrite
\ba
	\ket{\psi^\prime}&=\frac{1}{\sqrt{\bra{\psi}\mathcal{V}^\dagger \mathcal{V}\ket{\psi}}}\mathcal{V}\mathcal{U} c^{\dagger}_1\cdots c^{\dagger}_N\ket{0}\nonumber\\
    &=\frac{1}{\sqrt{\bra{\psi}\mathcal{V}^\dagger \mathcal{V}\ket{\psi}}}
	\prod_{k=1}^N(\mathcal{W}c^{\dagger}_k \mathcal{W}^{-1})\ket{0}
    \nonumber\\
    &=\frac{1}{\sqrt{\bra{\psi}\mathcal{V}^\dagger \mathcal{V}\ket{\psi}}}\prod_{k=1}^{N}\left[\sum_{j=1}^{L} w_{j k} c_{j}^{\dagger}\right]\ket{0}\,,
\end{align}
where $\mathcal{W}=\mathcal{V}\mathcal{U}$\,. Next, using the QR-decomposition
\begin{equation}
	w=qr\,,
\end{equation}
[$r$ is an upper triangular matrix] we have
\ba
	\ket{\psi^\prime}&=\frac{1}{\sqrt{\bra{\psi}\mathcal{V}^\dagger \mathcal{V}\ket{\psi}}}\mathcal{Q}\prod_{k=1}^{N}\left[\sum_{j=1}^{L} r_{j k} c_{j}^{\dagger}\right]\ket{0} \nonumber\\
    &=\frac{(\prod_{k=1}^{N}r_{kk})}{\sqrt{\bra{\psi}\mathcal{V}^\dagger \mathcal{V}\ket{\psi}}}\mathcal{Q}c_{1}^{\dagger}\ldots c_{N}^{\dagger}\ket{0}\,,
\end{align}
where we used $c^{\dagger}_kc^{\dagger}_k=0$, and the fact that $r$ is upper triangular.

Now, because $\mathcal{Q}$ is unitary by construction, $\mathcal{Q}c_{1}^{\dagger}\ldots c_{N}^{\dagger}\ket{0}$ is normalized. Thus (up to an irrelevant phase)
\begin{equation} 
	\prod_{k=1}^{N}r_{kk}=\sqrt{\bra{\psi}\mathcal{V}^\dagger \mathcal{V}\ket{\psi}}\,,
\end{equation}
and so
\begin{equation}
	\ket{\psi^\prime}=\prod_{k=1}^{N}\left[\sum_{j=1}^{L} q_{j k} c_{j}^{\dagger}\right]\ket{0}\,,
\end{equation}
with covariance matrix
\begin{equation}\label{eq:new_covariance}
	\braket{\psi^\prime|c^{\dagger}_jc_k|\psi^\prime}=q
	\begin{pmatrix}
		\openone_N & 0 \\
		0 & 0 
	\end{pmatrix} q^\dagger\,.
\end{equation}
This proves the validity of the QR decomposition used in Ref.~\cite{cao2019entanglement}. Importantly, just like the method developed in Ref.~\cite{bravyi2004lagrangian}, the QR decomposition is exact, and provides a physical covariance matrix not just at the first order in the Trotterization step, making it suitable for numerical implementation.

Now, suppose the initial state is not of the form~\eqref{eq:dw_initial_state}. In our work, we consider in particular the Néel state
\begin{equation}
    \ket{\psi_{\rm N}}=c^\dagger_1c^\dagger_3\ldots c^\dagger_{2N-1}\ket{0}\,.
\end{equation}
In this case, we can write
\begin{equation}\label{eq:state_neel}
    \ket{\psi_{\rm N}}=\mathcal{S}_0 c^{\dagger}_1 c^{\dagger}_2 \cdots c^{\dagger}_N\ket{0}\,,
\end{equation}
where $\mathcal{S}_0$ is a Gaussian unitary operator. The single-particle-sector matrix $s_0$ corresponding to $\mathcal{S}_0$ is simply a permutation matrix mapping the first $N=L/2$ sites into the odd sites of the chain. The representation~\eqref{eq:state_neel} allows us to repeat the previous algorithm, viewing $\mathcal{S}_0$ as implementing an initial unitary evolution.

\section{Further numerical data}
\label{app:more-numerics}

\begin{figure*}
    \centering
    \includegraphics[width=0.47\linewidth]{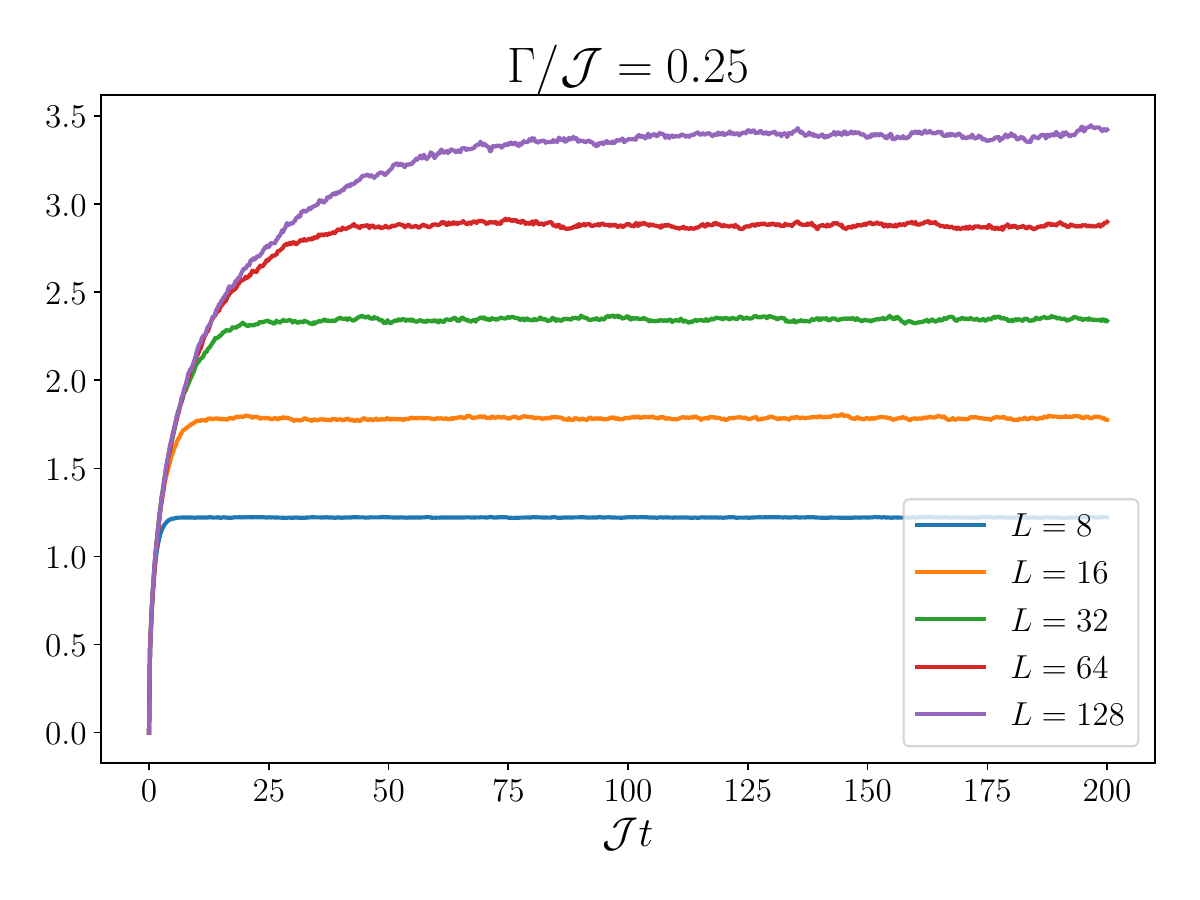}
    \includegraphics[width=0.47\linewidth]{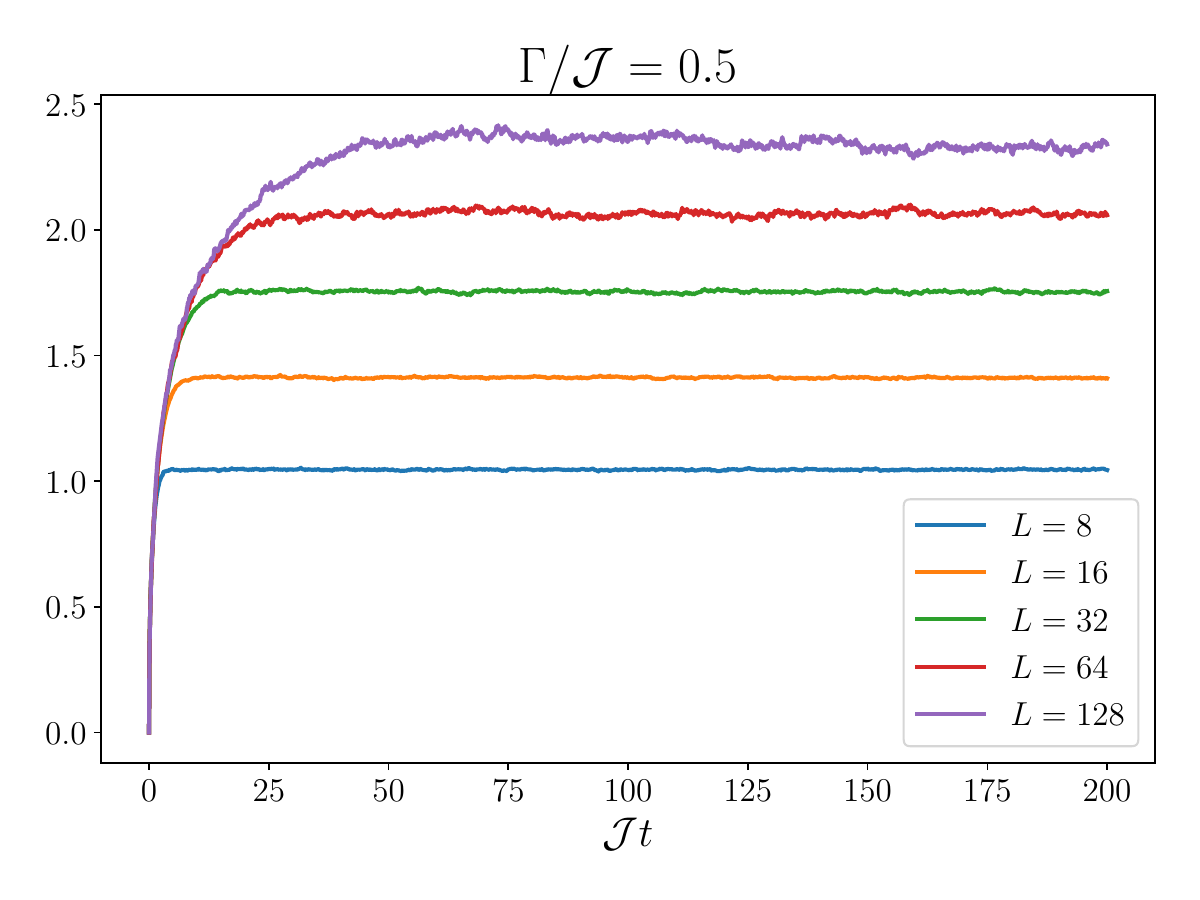}\\
    \includegraphics[width=0.47\linewidth]{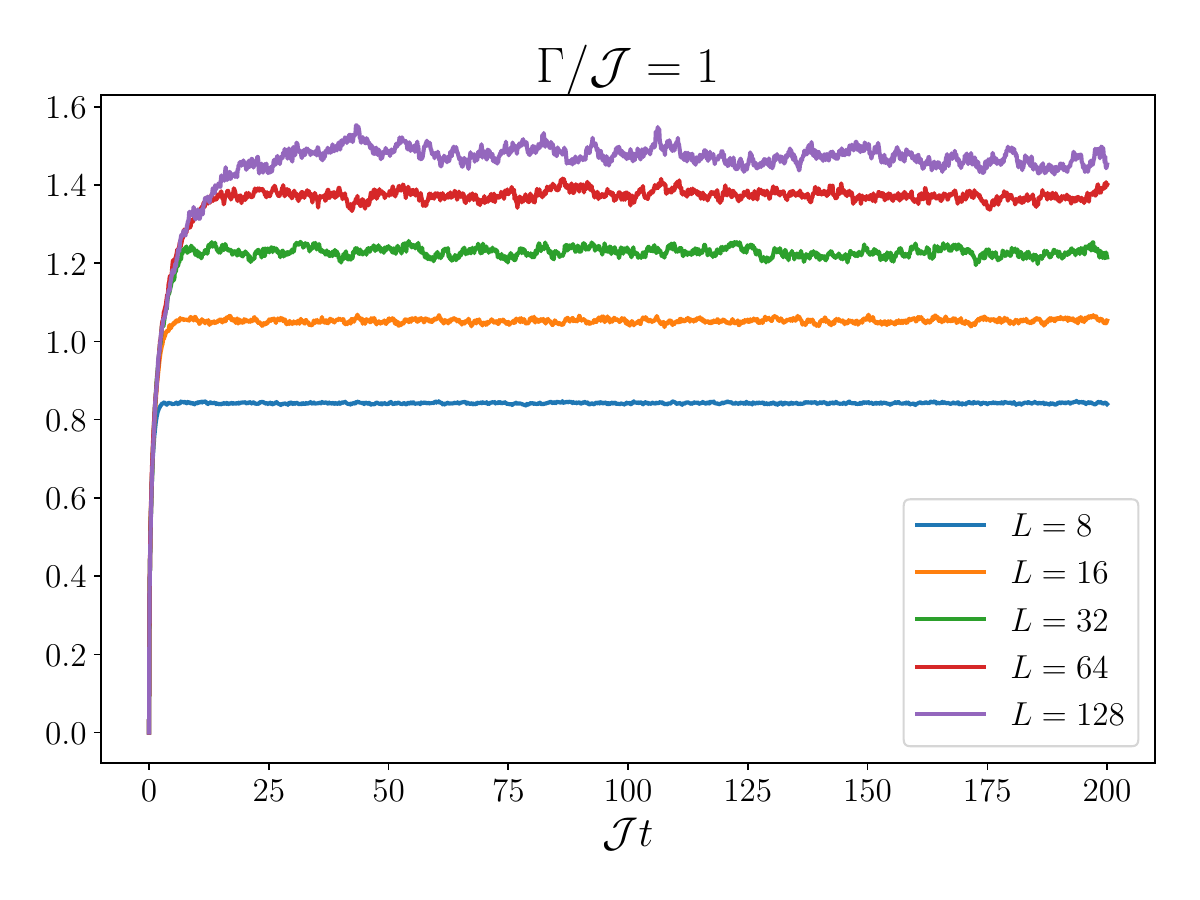}
    \includegraphics[width=0.47\linewidth]{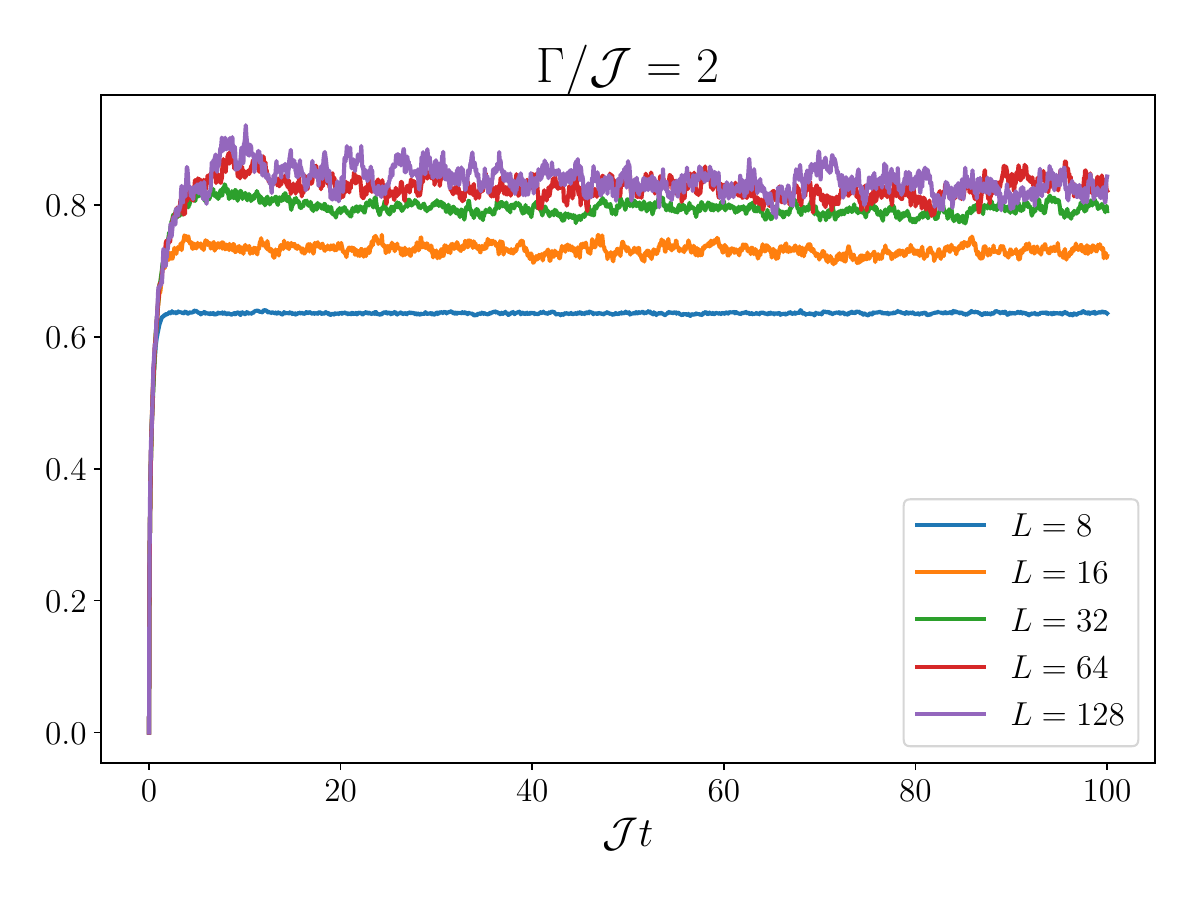}
    \caption{Average von Neumann entanglement entropy of half the system $S_{\rm vN}$ as a function of time $t$ and for various system sizes $L$. The number of samples used for averaging depends on the plot but is at least 80 for all plots.}
    \label{fig:S-vs-t}
\end{figure*}

In this Appendix we report more details concerning the numerical data reported in Fig.~\ref{fig:plateaux} and~\ref{fig:slope}.
The average values as a function of time are reported in Fig.~\ref{fig:S-vs-t}.
To compute the plateaux value, we averaged $S_{\rm vN}(t)$ starting from time $t_{\rm plateau} = 130,\,90,\,50,\,50$ for $\Gamma/\mJ = 0.25,\,0.5,\,1,\,2$ respectively. We checked that significantly increasing $t_{\rm plateau}$ changes the average value of $S_{\rm vN}$ by an amount compatible with the associated statistical error.

The simulations are performed using respectively $\delta t = 0.02,\,0.01,\,0.01,\,0.005$ for the different values of $\Gamma$ listed above. We checked that increasing $\delta t$ by a factor of $2$ did not significantly affect the plateau value of $S_{\rm vN}$.

\section{Extended symmetries for fixed-phase hopping}
\label{appendix:extended-sym}

The presence of the extended symmetry is most easily shown in the case of imaginary hopping amplitudes ${J_j \in i \mathbb{R}}$.
(It does not matter whether these depend on time.)
Redefining the fermionic operators via a particle-hole transformation on the $\sigma=-$ replicas,
\be
    b_j^{(\sigma a)} =
    \begin{cases}
        c_j^{(\sigma a)} & \text{if }\sigma=+,\\
        c_j^{(\sigma a)\,\dag} & \text{if }\sigma=-.\\
    \end{cases}
\ee
and expressing the replicated Hamiltonian~\eqref{eq:replicated-Hamiltonian} in terms of these, it is immediate to verify that the Hamiltonian is invariant under
\be
    b_j^{\alpha} \mapsto U^{\alpha \beta} b_j^{\beta}
\ee
for any $\mathrm{U}(2N)$ unitary matrix $U$. 

This consideration applies to any lattice (bipartite or not bipartite) with imaginary hoppings, including regular lattices in any number of dimensions, 
as  the replicated Hamiltonian~\eqref{eq:replicated-Hamiltonian} has a similar form on an arbitrary lattice.

Having established the extended symmetry for imaginary hoppings, the more general case discussed in the main text follows by making a simple change of gauge for the \textit{physical} fermion operators, ${c_j^\dag \rightarrow e^{i\chi_j} c_j^\dag}$. If it is possible to find a gauge where all the hoppings are imaginary then the extended symmetry holds. For example, this is always possible in a nearest-neighbor chain (with $N_F=1$), if the phases of the hoppings are time-independent.

Finally, we give a heuristic argument for the manifold of the \NLSM for models which obey the constraint discussed in this appendix.

For a concrete example, consider the model of the main text with $N_F=1$ but with imaginary random $J_i(t)$.
We assume that the semiclassical ground states may be characterized by expectation values of fermion bilinears (as in the case described in the main text) and that there is a semiclassical ground state that breaks replica symmetry in the same way as the state $\ket{\mathbb{I}}$ defined in App.~\ref{sec:boundar_states}.
Recall that this state appears in the computation of traces and entropies and defines a choice of pairing of forward and backward replicas. 

After averaging, and the particle-hole transformation described above, the effective Hamiltonian may be written in  terms of the $4N\times 4N$ matrix of fermion bilinears 
\be
A = \begin{pmatrix}
b^{(+)}\\  
b^{(-)}\\
b^{(+)\dag}\\
b^{(-)\dag}
\end{pmatrix}
\begin{pmatrix}
b^{(+)\dag},  
b^{(-)\dag},
b^{(+)},
b^{(-)}
\end{pmatrix}
- \f{1}{2} \mathds{1}_{4N},
\ee
where e.g. $b^{(+)}$ is an $N$-component vector indexed by a replica label $a=1,\ldots, N$.
The expectation value in the state $\ket{\mathbb{I}}$ 
is 
\be\label{eq:4Nby4Ndecomp}
A = \begin{pmatrix}
{0}_{2N} &  \f{i}{2} B \\
 -\f{i}{2}B^* & {0}_{2N}
\end{pmatrix},
\ee
with 
\be
B = \begin{pmatrix}
0_N&  -\mathds{1}_N \\
\mathds{1}_N & {0}_{N}
\end{pmatrix}.
\ee
SU$(2N)$ rotations of the vector $(b^{(+)}, b^{(-)})$ preserve the form in (\ref{eq:4Nby4Ndecomp}), but rotate $B$ into
\be
B \mapsto U_{2N} \begin{pmatrix}
0_N&  -\mathds{1}_N \\
\mathds{1}_N & {0}_{N}
\end{pmatrix} 
U^\top_{2N}.
\ee
The subgroup of SU$(2N)$ that leaves $B$ invariant is therefore by definition $\mathrm{Sp}(2N)$, 
sometimes denoted $\mathrm{USp}(2N)$.
Therefore the orbit of $B$ is the manifold $\mathrm{SU}(2N)/\mathrm{Sp}(2N)$.

\bibliography{refs}

\end{document}